**Surface ferromagnetism in rhombohedral heptalayer graphene moiré superlattice**

Wenqiang Zhou[1,2#], Jing Ding[1,2#], Jiannan Hua[1,2#], Le Zhang[1,2], Kenji Watanabe[3], Takashi Taniguchi[4], Wei Zhu[1,2*], Shuigang Xu[1,2*]

[1] *Key Laboratory for Quantum Materials of Zhejiang Province, Department of Physics, School of Science, Westlake University, 18 Shilongshan Road, Hangzhou 310024, Zhejiang Province, China*

[2] *Institute of Natural Sciences, Westlake Institute for Advanced Study, 18 Shilongshan Road, Hangzhou 310024, Zhejiang Province, China*

[3] *Research Center for Electronic and Optical Materials, National Institute for Materials Science, 1-1 Namiki, Tsukuba 305-0044, Japan*

[4] *Research Center for Materials Nanoarchitectonics, National Institute for Materials Science, 1-1 Namiki, Tsukuba 305-0044, Japan*

[#]These authors contributed equally to this work.

[*]Correspondence to: zhuwei@westlake.edu.cn, xushuigang@westlake.edu.cn

**The topological electronic structure of crystalline materials often gives rise to intriguing surface states, such as Dirac surface states in topological insulators[1], Fermi arc surface states in Dirac semimetals[2], and topological superconductivity in iron-based superconductors[3]. Recently, rhombohedral multilayer graphene has emerged as a promising platform for exploring exotic surface states due to its hosting of topologically protected surface flat bands at low energy, with the layer-dependent energy dispersion[4-8]. These flat bands can promote electron correlations, leading to a plethora of quantum phenomena, including spontaneous symmetry breaking[9], superconductivity[10-12], ferromagnetism[13,14], and topological Chern insulators[15-17]. Nevertheless, the intricate connection between the surface flat bands in rhombohedral multilayer graphene and the highly dispersive high-energy bands hinders the exploration of correlated surface states. Here, we present a method to isolate the surface flat bands of rhombohedral heptalayer (7L) graphene by introducing moiré superlattices. The pronounced screening effects observed in the moiré potential-modulated rhombohedral 7L graphene indicate its essential three-dimensional (3D) nature. The isolated surface flat bands favor correlated states on the surface in the regions away from charge-neutrality points. Most notably, we observe tunable surface ferromagnetism, manifested as an anomalous Hall effect with hysteresis loops, which is achieved by polarizing surface states using finite displacement fields. Our work establishes rhombohedral multilayer graphene moiré superlattice as a unique 3D system for exploring correlated surface states.**

In rhombohedral multilayer graphene, the low-energy electrons are entirely concentrated on the two surface layers, while the bulk states exhibit an energy gap[18]. This distinctive characteristic provides an ideal platform for the exploration of diverse surface states. The surface electronic bands of rhombohedral graphene can be approximately described by $E \sim \pm p^N$ in a two-band model, where $E$ is the kinetic energy, $p$ the momentum, and $N$ the layer number[18,19]. With increasing $N$, these surface bands become extremely flat at low energy. Due to the instability to electronic interactions endowed by their large density of states (DOS), these surface flat bands hypothetically host strongly correlated states, such as spontaneous quantum Hall states[20], high-temperature superconductivity[21], ferromagnetism[22-25]. Furthermore, in rhombohedral multilayer graphene, the low-energy surface states, characterized by alternating intralayer and interlayer hopping, is a tailor-made simulator of one-dimensional topological insulator in the Su-Schrieffer-Heeger model[26]. Moreover, the chiral stacking in rhombohedral graphene gives rise to large momentum-space Berry curvatures, accompanied by a giant intrinsic magnetic moment inherited from the multivalley features of graphene. This feature positions it as a promising platform for exploring topological non-trivial states, such as anomalous Hall effect (AHE)[5]. Experimentally, with recent advancement of technique for producing hexagonal boron nitride (h-BN) encapsulated structures[27], correlation-driven insulating states, magnetic states, and superconductivity have been reported in bilayer ($N = 2$)[28], rhombohedral trilayer ($N = 3$)[7,8,29-31], tetralayer[32], pentalayer[33,34], and multilayers ($N \geq 7$) graphene[6,35].

The power-law energy dispersion in rhombohedral multilayer graphene suggests that the low-energy surface flat bands are connected to highly dispersive high-energy bands. Consequently, the observations of strong correlations in intrinsic rhombohedral graphene have been restricted to very low carrier density ($n$) regimes ($n \to 0$)[6-8]. To isolate these surface flat bands from the high-energy dispersive bands is not only beneficial for exploring correlated states in high $n$ regimes, but also indispensable for recurring Chern band. One versatile approach for achieving this isolation is through the stacking of van der Waals materials with a twist angle and/or a lattice mismatch, which constructs moiré superlattices at two-dimensional (2D) interfaces[36,37]. These moiré superlattices impose a long-range periodic potential, resulting in band folding and the formation of a mini-Brillouin zone. This process typically leads to bandwidth reduction, thereby enhancing the effects of electronic correlations. Consequently, many unique band structures emerge at low energy near the Fermi surface, accompanied by the appearance of exotic states, such as superconductivity[10], correlated insulating states[38], orbital magnetism[13], and Hofstadter butterfly[39,40].

Here, we introduce moiré superlattices into rhombohedral multilayer graphene, to separate the low-energy surface flat bands away from high-energy dispersive bands. These moiré superlattices were constructed by crystallographically aligning rhombohedral multilayer graphene with h-BN during the van der Waal assembly. Thanks to the small lattice mismatch ($\delta \approx 1.6\%$) between graphene and h-BN, a moiré superlattice can be formed with a long-range wavelength given by $\lambda = \frac{(1+\delta)a_G}{\sqrt{2(1+\delta)(1-\cos\theta)+\delta^2}}$, where $a_G = 0.246$ nm is the in-plane lattice constant of graphite, and $\theta$ the relative misalignment angle between the two lattices. Our band calculations confirm the presence of an isolated surface flat band at the conduction band, as shown in Fig. 1e and Extended Data Fig. 12. To probe the electronic transport of rhombohedral graphene, we have employed a dual-gate structure, as depicted schematically in Fig. 1b, which enables us to independently control $n$ and displacement field ($D$).

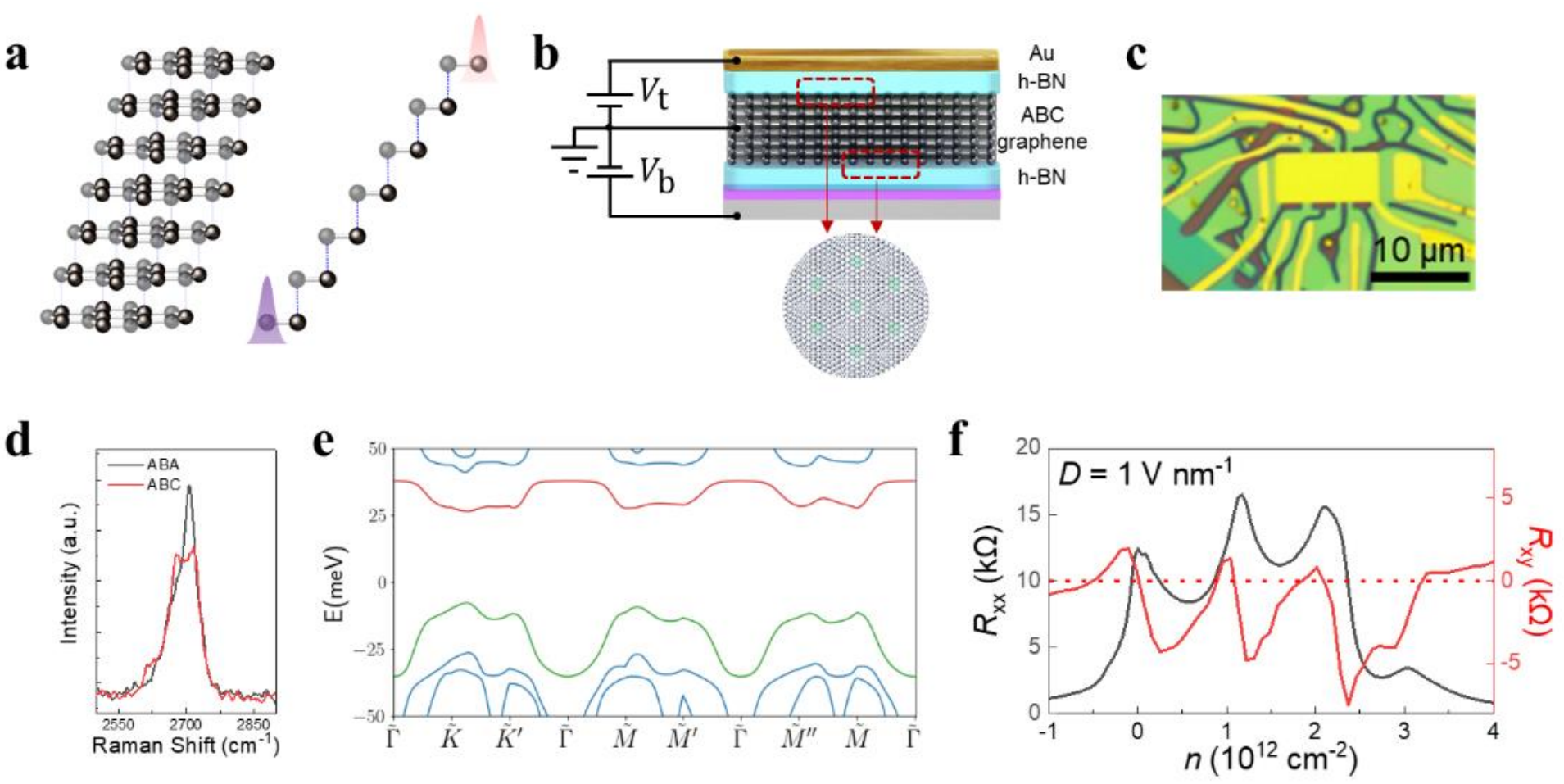

**Fig. 1| Rhombohedral 7L graphene moiré superlattice**. **a**, Schematic of rhombohedral 7L graphene. Left and right represent side view and cross-section view along the in-plane armchair direction, respectively. The two curves in the right schematic illustrate the wavefunctions of low-energy states concentrate at the sublattices located at each surface. **b**, Schematic of a dual-gate h-BN encapsulated devices with moiré superlattices at the interfaces between h-BN and graphene. **c**, Optical image of a typical device with a Hall bar geometry. **d**, Raman spectra of ABA-stacked and ABC-stacked 7L graphene. **e**, Calculated band structure of rhombohedral 7L graphene with moiré superlattice at both top and bottom surface. The interlayer potential used in the calculation is 12 meV. **f**, Longitudinal ($R_{xx}$) and Hall ($R_{xy}$) resistances as a function of total carrier density measured at $D = 1$ V nm$^{-1}$ and $T = 50$ mK.

Our devices were fabricated through the mechanical exfoliation of natural graphite. We chose rhombohedral heptalayer (7L) graphene as the building block, since our previous work indicates that it preserves the three-dimensional (3D) characteristics of graphite meanwhile exhibiting strong correlations[6]. Raman spectra and mapping techniques were employed to identify the stacking order and select rhombohedral (also described as ABC) domains for device fabrication (see Fig. 1d and Extended Data Fig. 2). Fig. 1f shows low-temperature ($T = 50$ mK) longitudinal ($R_{xx}$) and Hall ($R_{xy}$) resistances as a function of $n$, with carriers concentrated at one of the surfaces under a fixed $D = 1$ V nm$^{-1}$. Besides the peak at charge-neutrality point ($n = 0$), $R_{xx}$ exhibits two additional prominent peaks at high-density region. The corresponding $R_{xy}$ exhibits sign reversals, indicative of Fermi surface reconstruction[41]. This phenomenon can be attributed to either band folding caused by the moiré superlattice or strong correlations, which we will discuss in detail later. In either case, with the assistance of the moiré superlattice, we have succeeded in isolating the surface band from high-energy band, resulting in the opening of a band gap in high $n$ regions.

To reveal the electronic transport behavior influenced by the moiré potential in rhombohedral 7L graphene, we also fabricated a device using intrinsic rhombohedral 7L graphene without alignment with h-BN (device D1) for reference. Fig. 2a and 2b show color maps of $R_{xx}(n, D)$ for devices without and with moiré superlattice, respectively. In the absence of moiré, two distinct insulating states emerge at $n = 0, D = 0$, and $n = 0, |D| > 0.4$ V nm$^{-1}$ as illustrated in Fig. 2a. This behavior closely resembles what has been observed in rhombohedral nonalayer (9L) graphene[6]. The insulating state at $|D| > 0.4$ V nm$^{-1}$ is attributed to the opening of an energy gap in the surface states, resulting from inversion symmetry breaking induced by a large electric field. Differently, the insulating state at $n = 0, D = 0$ cannot be explained in a single-particle picture and is believed to be a correlated gap as a result of spontaneous symmetry breaking favored by surface flat band[20]. It's noted that the insulating states at $n = 0, D = 0$ strongly rely on the electronic coupling between top and bottom surfaces, which only occurs in thin-layer (roughly $N \leq 10$) rhombohedral graphene[6]. In rhombohedral 7L graphene, this correlated gap is highly reproducible and has been observed in multiple devices (see Extended Data Fig. 7).

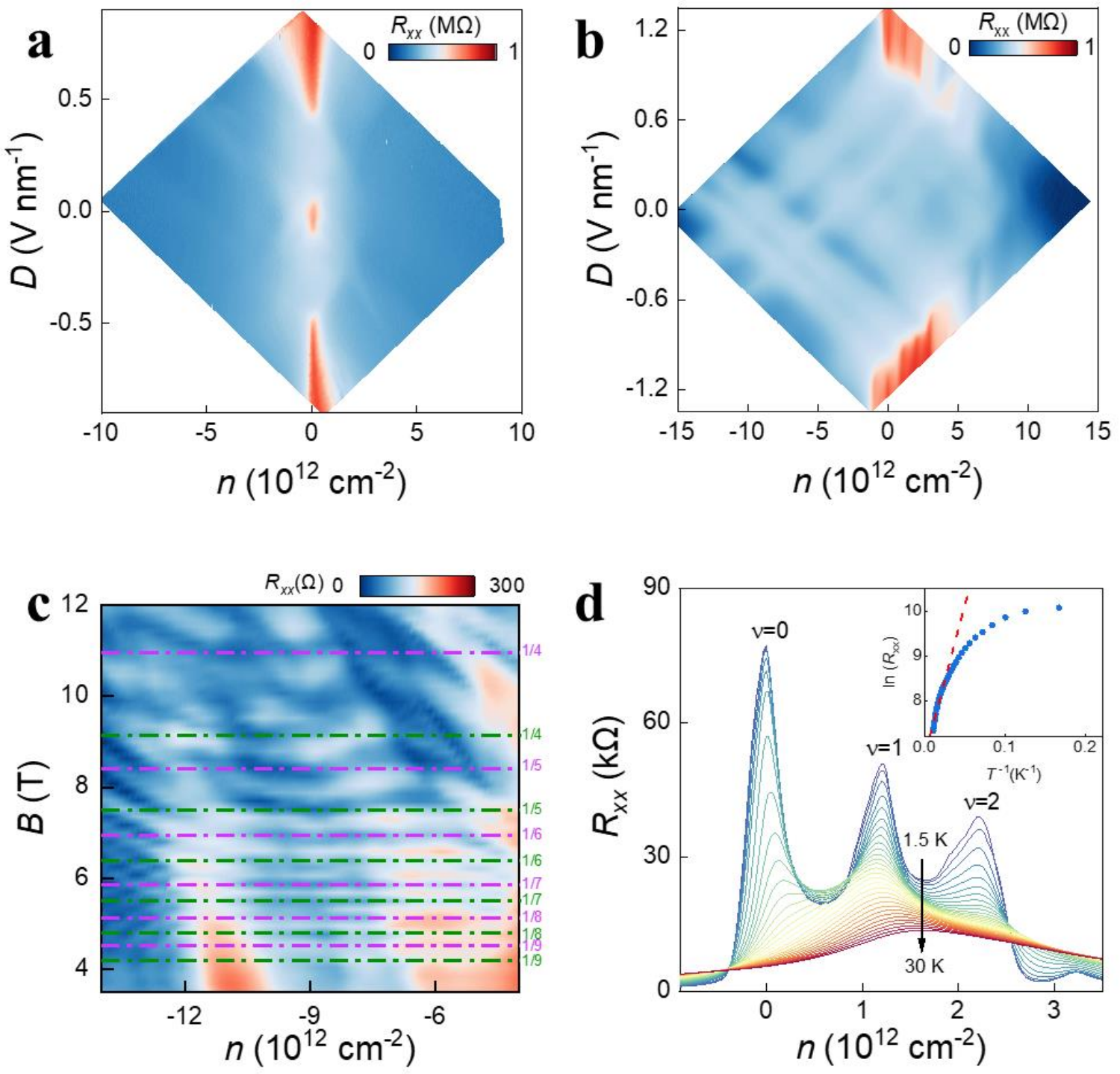

**Fig. 2| Low-temperature transport characteristics of rhombohedral 7L graphene without and with moiré superlattice**. **a, b**, Color maps of longitudinal resistance $R_{xx}$ plotted in logarithmic scales as a function of carrier density $n$ and displacement field $D$ measured at $T = 50$ mK and $B = 0$ T for the devices without (**a,** device D1) and with (**b,** device D2) moiré superlattice. **c**, $R_{xx}$ as a function of magnetic field $B$ and total carrier density $n$ at $T$ =50 mK and $D = 0$ V nm$^{-1}$. Quantum oscillations independent on $n$ were observed. We attribute it as Brown-Zak oscillations arising from moiré potentials at the two interfaces between graphene and h-BN. There are two sets of oscillatory, indicating this sample is doubly aligned with two decoupled moiré superlattices at each interface. The labels near y axis denotes the $\frac{\phi}{\phi_0} = 1/q$, when the integer number $q$ of superlattice unit cells are commensurate with the magnetic flux quantum $\phi_0$. **d**, Temperature dependence of $R_{xx}$ as a function of $n$ at $D = 1.1$ V nm$^{-1}$, $B = 0$ T. Inset: Arrhenius plot (ln$R_{xx}$ versus $T^{-1}$) for charge-neutrality point ($\nu = 0$) at high temperature region. The dashed line represents the linear fit, from which the transport gaps $\Delta$ can be extracted by $\ln R_{xx} \propto \Delta/2k_BT$. The linear fits give $\Delta = 12.9$ meV, 4.7 meV, and 0.8 meV at $\nu = 0, 1,$ and $2$, respectively. The data in **(c)** and **(d)** were acquired in the sample with moiré superlattice (device D2).

Introducing moiré superlattice into rhombohedral 7L graphene significantly modifies its transport behavior, as shown in Fig. 2b (device D2). First, the correlated gap at $n = 0, D = 0$ disappears, indicating moiré potential at the interface between h-BN and graphene effectively decouples the two surface states. Second, the critical field ($D_c$), above which a band insulator gap is opened, increases to approximately 0.8 V/nm. Applying $D$ via asymmetric dual gates generates a potential difference between the two surfaces, resulting in a carrier redistribution that strongly screens out the external field. The larger $D_c$ in Fig. 2b indicates the moiré potential favors carrier localization at the surfaces, thus enhancing the screening effect. This enhanced screening effect is further evident from the presence of a series of horizontal and vertical lines in the region below $D_c$, when plotting $R_{xx}$ as a function of $n_t$ and $n_b$ (see Extended Data Fig. 4). It serves to electronically decouple the two surface states and suppress their interactions, which explains the absence of correlated states at $n = 0, D = 0$. These features collectively indicate that moiré potential modified-rhombohedral 7L graphene essentially behaves as a 3D system.

Third, we also observed additional gap states at large $D$ beyond charge-neutrality point ($n \neq 0$). When $|D| > |D_c|$, the finite band overlap between conduction and valence bands is lifted due to inversion symmetry breaking. The surface states become fully polarized, such that charge carriers concentrate on only one of the two surfaces. Namely, for positive $D > D_c$, only electrons (holes) in the conduction (valence) band at the bottom (top) surface contribute to the conductance (see Extended Data Fig. 4). The screening effect vanishes, manifested as both gates are effective, accompanied by the disappearance of the horizontal and vertical lines in Extended Data Fig. 4a. Unlike device D1 in Fig. 2a, device D2 exhibits additional resistance peaks at $n_1 = 1.0 \times 10^{12}$ cm$^{-2}$ and $n_2 = 2.1 \times 10^{12}$ cm$^{-2}$ for $D > 0$. Similar extra prominent peaks also appear for $D < 0$, but at slightly different $n$. Notably, when comparing these features to those in device D1 without moiré, the peaks appearing at non-zero $n$ stem from the formation of moiré minibands. The observation of remarkable Brown-Zak oscillations[38,40,42], as shown in Fig. 2c, further confirms the formation of moiré superlattices in Device D2. In Fig. 2c, there are two distinct sets of oscillatory behavior periodic in $1/B$, which indicates that our device has doubly aligned configuration[43,44]. From the oscillation period, we can extract two twist angles $\theta_1 = 0.88°$ and $\theta_2 = 0.90°$ at the two interfaces. With this, we can assign $n_1$ and $n_2$ to the quarter filling ($\nu = 1$) and half filling ($\nu = 2$) of the moiré miniband, respectively.

The double alignment is consistent with that two sets of extra peaks at $n > 0$ appear at both $D > 0$ and $D < 0$. The possibility of double alignment can be further confirmed by the optical image of the stack showing the alignment of the straight edges of h-BN and graphene flakes (see Extended Data Fig. 3). As our graphene is sufficiently thicker, the moiré superlattices from the two interfaces remain decoupled, a fact supported by the aforementioned several features. It's reasonable to treat the moiré effects as independence and disregard the super-moiré effects. In the following part, we mainly focus on the $n > 0, D > 0$ region, namely on the conduction band modulated by the moiré superlattice at the bottom surface. Similar behavior can be found in the other regions (see Extended Data Fig. 10).

The temperature dependence of resistance peaks at $\nu = 1$ and $\nu = 2$ exhibits typical insulating behavior, where the resistance increases as the temperature decreases. These insulating states at partial fillings are correlated insulators, arising from strong electron-electron interactions induced spontaneous symmetry breaking, facilitated by the further flattening of surface bands through zone folding. From the Arrhenius plots in Fig. 2d and Extended Data Fig. 8, we estimate that the single particle gap and correlated gaps at $\nu = 1$ and $\nu = 2$ are approximately 12.9 meV, 4.7 meV, 0.8 meV, respectively.

One remarkable feature of rhombohedral 7L graphene, compared with thinner one, lies in its 3D characteristic, which can be further evident from Landau quantization at high $B$. Fig. 3 shows the Landau diagrams at $B = 14$ T of rhombohedral 7L graphene both without and with moiré superlattices. In both devices, a series of horizontal and vertical features emerge at small $D$, which arise from the coexistence of two surface states with a high carrier density, effectively screening out their influence on each other. As $D$ increases above a critical value, the surface states become polarized, with carriers concentrating on only one of the two surfaces. In this regime, the carriers can be effectively tuned by both gates, resulting in the appearance of diagonal Landau levels (LLs) features at high $D$. The prominent screening features observed in Landau diagrams resemble those seen in Bernal (ABA) -stacked graphite, albeit with opposite distributions[45-47] (see detailed discussions in Methods).

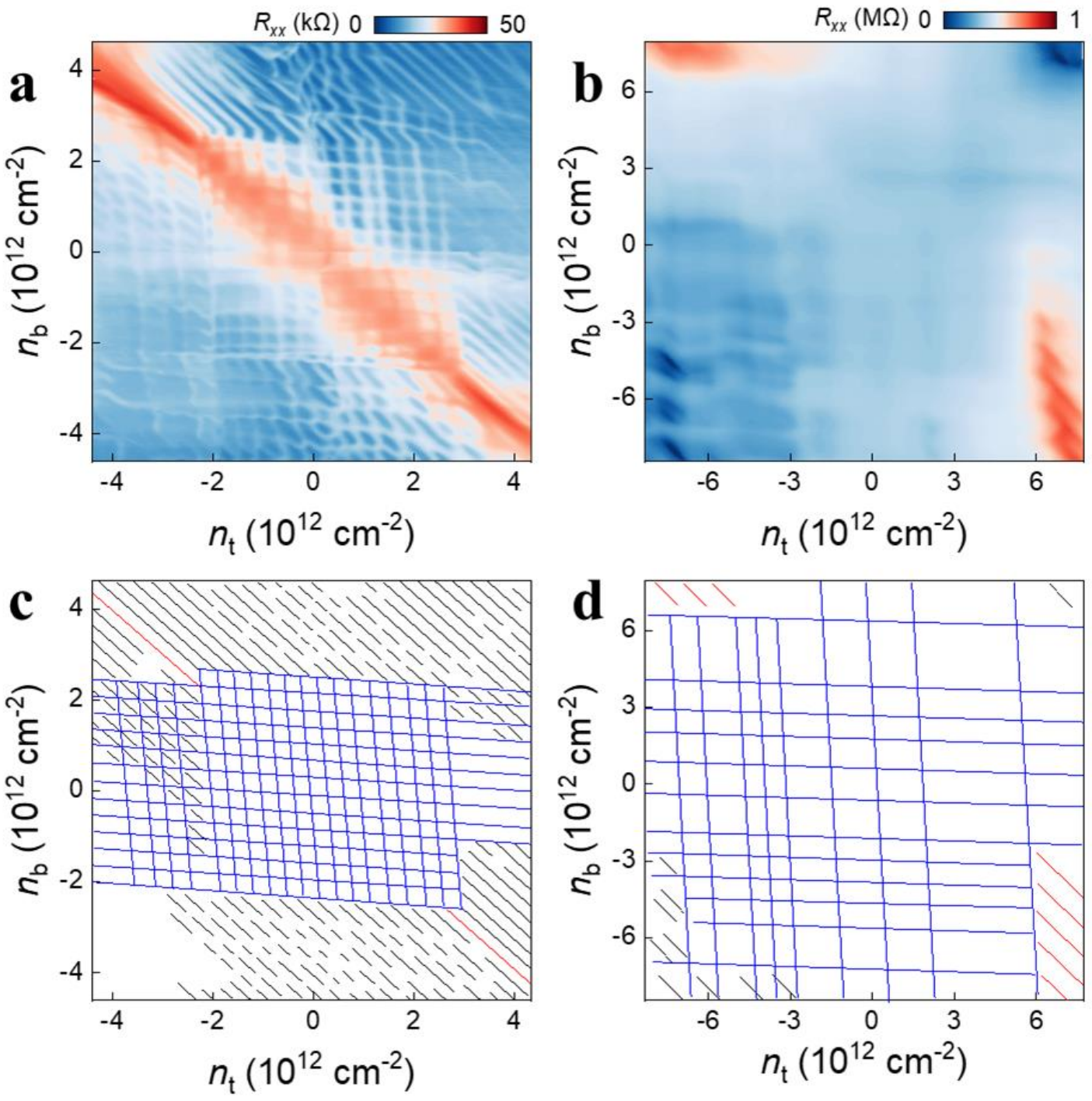

**Fig. 3| Landau quantization in rhombohedral 7L graphene. a, b**, Color maps of longitudinal resistance $R_{xx}$ as a function of top and bottom gates induced carrier density $n_t$ and $n_b$, measured at $B = 14$ T. The data were taken from the devices without (**a,** device D3 at $T = 50$ mK) and with (**b,** device D2 at $T = 1.5$ K) moiré superlattice. **c**, **d**, Wannier diagram depicting the LLs according to the raw data in **a** and **b**, respectively. The red lines represent the insulating states at zero field. The black lines represent LLs from polarized surface states, manifested as diagonal lines tunable by both gates. The blue lines represent screened LLs, where two surface states are strongly screened by each other, manifested as a series of horizontal and vertical features.

Further insights into the broken symmetry can be gained from the Hall resistance $R_{xy}$. Fig. 4a and 4b provide high-resolution maps of $R_{xx}$ and $R_{xy}$ in the vicinity of the correlated insulating states at $D > 0$. Near $\nu = 1$ and $\nu = 2$, maxima in $R_{xx}$ are accompanied with rapid sign changes in $R_{xy}$, indicating a change of carrier type. These sign reversals result from Fermi surface restructuring driven by correlations and the formation of a new band edge, similar to that in twisted graphene[41]. Additionally, the evolution of $R_{xy}$ as a function of $n$ and $D$ at low $B$ in Fig. 4b exhibits gradual sign changes within the $n$ filling from $\nu = 0$ to $\nu = 1$ and $\nu = 1$ to $\nu = 2$. These sign changes correspond to divergences in $n$ and are associated with saddle points in the energy dispersion at the Fermi surface, known as van Hove singularities (vHSs) [48].

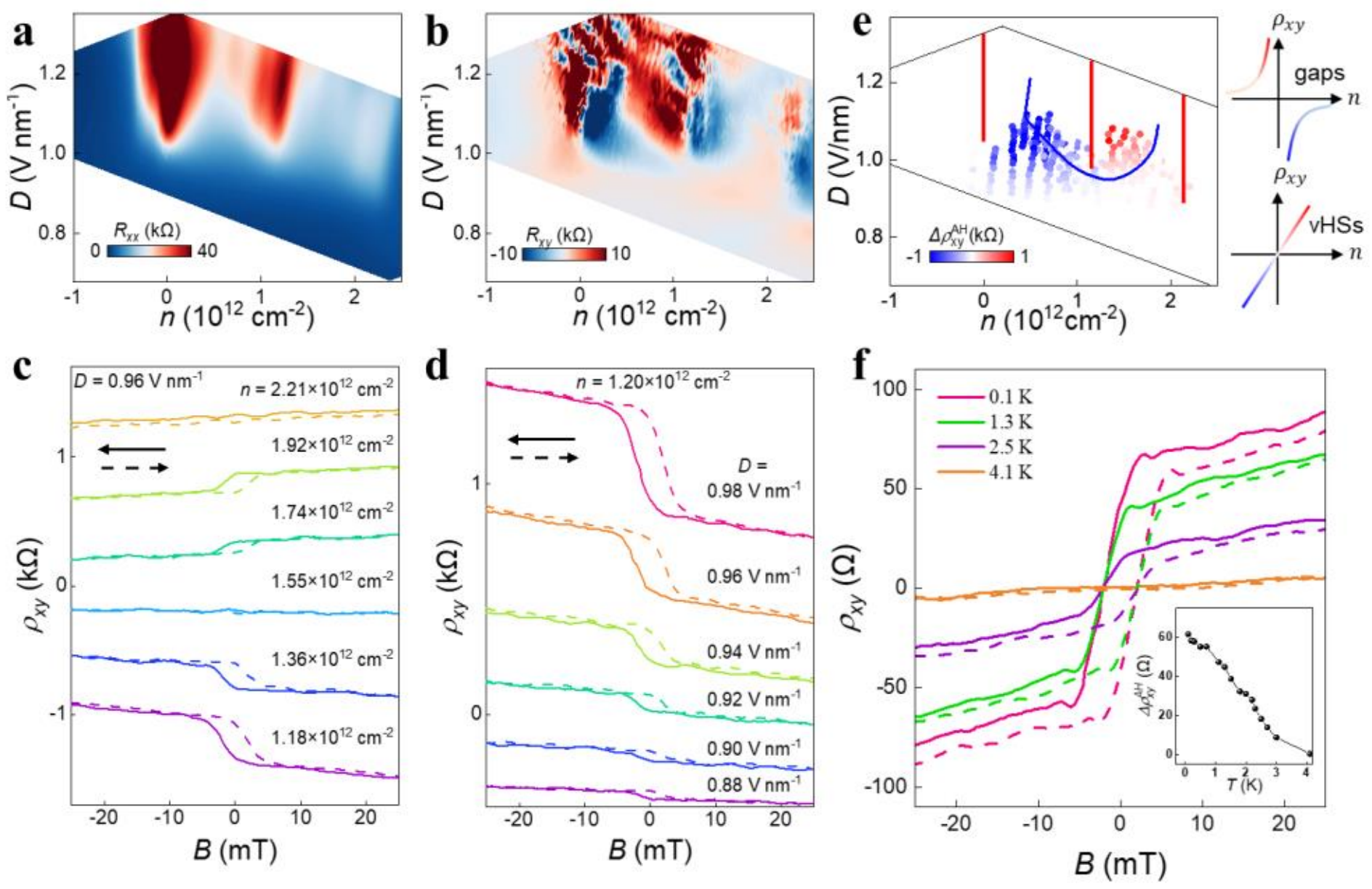

**Fig. 4| Tunable surface ferromagnetism in rhombohedral 7L graphene moiré superlattice. a,** Fine maps of $R_{xx}$ plotted in linear scales as a function of $n$ and $D$ near the conduction band modulated by the moiré superlattice at the bottom surface for $B = 0$. **b**, Corresponding anti-symmetrized Hall resistance $R_{xy} = \frac{R_{xy}(+B)-R_{xy}(-B)}{2}$ at a fixed small magnetic field $B = \pm 1$ T. The Hall resistance changes its sign at gap states and vHSs with different ways. **c, d**, Anti-symmetrized Hall resistance $\rho_{xy}$, defined in the Methods, as a function of $B$ swept back and forth at (**c**) a fixed $D = 0.96$ V nm$^{-1}$, varying $n$ from $1.18 \times 10^{12}$ cm$^{-2}$ to $2.12 \times 10^{12}$ cm$^{-2}$, and (**d**) a fixed $n = 1.20 \times 10^{12}$ cm$^{-2}$, varying $D$ from 0.88 V nm$^{-1}$ to 0.98 V nm$^{-1}$. The absolute values are manually offset for clarity. The AHE with both nonlinear features and hysteresis loops manifests itself as a ferromagnetic state, which is tunable by both $n$ and $D$. **e**, Color plots of residual resistance $\Delta\rho_{xy}^{AH}$ as a function of $n$ and $D$. The individual dots were extracted from the measurements of AHE at corresponding $n$ and $D$. The colors represent the values of $\Delta\rho_{xy}^{AH}$. The red and blue curves denote two types of sign reversal of $R_{xy}$ near gaps and vHSs, respectively, which are sketched in the right schematic. **f,** Temperature dependence of AHE. $\rho_{xy}$ as a function of $B$ swept back and forth, showing ferromagnetic hysteresis at different temperatures, with fixed $n = 2.10 \times 10^{12}$ cm$^{-2}$ and $D = 0.91$ V nm$^{-1}$. The inset shows the evolution of residual resistance $\Delta\rho_{xy}^{AH}$ as a function of temperature. All the data of **a-f** were taken in device D2 at $T = 50$ mK.

When the Fermi energy approaches vHSs, the large DOS may lead to Fermi-surface instabilities, potentially giving rise to various exotic phases, such as superconductivity, ferromagnetism, and charge density waves. One particular example is the ferromagnetic instability, governed by the Stoner criterion[49]: $UD_F > 1$, where $U$ is the Coulomb energy, $D_F$ the DOS at the Fermi energy. The highly tunable vHSs, as shown in Fig. 4b, allow us to observe the Stoner ferromagnetism. Fig. 4c and 4d display the $n$ and $D$-dependent anti-symmetrized Hall resistance $\rho_{xy}$ (see Methods) when sweeping the out-of-plane $B$ back and forth between -25 mT and 25 mT. At $n = 2.21 \times 10^{12}$ cm$^{-2}$ and $D = 0.96$ V nm$^{-1}$, $\rho_{xy}$ exhibits normal linear behavior and remains independent of the sweep direction. But within a large region, $\rho_{xy}$ displays a remarkable AHE accompanied by hysteresis loops. The hysteresis becomes narrower with increasing $B$ and vanishes above a coercive field of $B_c = 7$ mT. At $B = 0$, $\rho_{xy}$ shows a nonzero value with the sign depending on the sweep direction of $B$, indicating the presence of remanent magnetization in the sample. These series of features are the hallmark of ferromagnetism, stemming from spontaneous time-reversal symmetry breaking within this system. We note that the observed hysteresis here is different from that in our previous work on intrinsic rhombohedral graphite near $n = 0$ and $D = 0$, where the hysteresis origins from electronic phase inhomogeneities[6]. Whereas, in the present system, strong interactions and a large DOS within the low-energy surface flat band are responsible for the emergence of ferromagnetism. Furthermore, the hysteresis displays no Barkhausen jumps upon sweeping $B$, a phenomenon often seen in twisted graphene systems[13,14], indicating the cleanness of the graphene/h-BN moiré superlattice system.

The Hall signal comprises both a linear component originating from the normal Hall effect and an anomalous component arising from the magnetization. After subtracting the linear component, we plot the anomalous residual resistance $\Delta\rho_{xy}^{AH}$ as a function of $n$ and $D$, shown in Fig. 4e, which reflects the evolution of the remanent magnetization strength. The $\Delta\rho_{xy}^{AH}$ values, whether positive or negative, are marked by red and blue colors, respectively, with the intensity of the colors representing the magnitude of AHE. From Fig. 4e, obviously the AHE is highly tunable by $n$ and $D$, with the largest values appearing in the vicinity of vHSs.

Fig. 4f shows the temperature-dependent hysteresis loops for a p-type-like carrier. The hysteresis of $\rho_{xy}$ disappears above a critical temperature, which further confirms the phase transition from ferromagnetism to paramagnetism. The Curie temperature, defined by the onset of hysteresis, is 4 K at an optimized position.

The ferromagnetism observed in rhombohedral 7L graphene moiré superlattice differs from those previously reported in other graphene system[7,13,28,30]. First, our system exhibits a pronounced 3D nature as aforementioned. Ferromagnetism in our system occurs only when electrons are entirely localized at one of the surface layers by applying high $D$. Namely, ferromagnetism observed here arises from electron interactions within individual surface layer. We refer to this as surface ferromagnetism or layer-polarized ferromagnetism. Second, Stoner ferromagnetism other than Chern band governs the AHE observed in our system. On the one hand, the emergence of ferromagnetism instabilities in our system spans a wide range, including non-integer moiré band fillings, and is enhanced near vHSs within the flat moiré bands. In contrast, in twisted bilayer graphene, the observed ferromagnetism typically occurs in a narrow region near an insulator at integer filling[13]. On the other hand, the residual $\Delta\rho_{xy}^{AH}$ near $B = 0$ in our system is relatively small (a few hundred Ohm), far from the quantized value of $h/e^2$. Third, the ferromagnetism in our system is exclusive to the conduction band, which is consistent with the calculated band structure in Fig. 1e showing an extremely narrow isolated conduction band. This contrasts with the ferromagnetism observed in the valence band of rhombohedral trilayer moiré superlattice[30].

To summarize, our results promote the observation of ferromagnetism in graphene systems from 2D to 3D. The emergence of ferromagnetism at the surface states is facilitated by the presence of flat surface bands, favored by both band structures of intrinsic rhombohedral graphene and the moiré superlattice. This work establishes rhombohedral multilayer graphene as a fertile platform for exploring novel surface states. The surface flat band in rhombohedral multilayer graphene moiré systems, when interplaying with nontrivial topological electronic states, may give rise to exotic correlated and topological physics, such as surface superconductivity[21] and quantum anomalous Hall effect in 3D system. The tunability of the layer number in rhombohedral graphene provides great potential for observing amazing quantum states. For example, during the preparation of this manuscript, the observation of fractional quantum anomalous Hall effect in rhombohedral pentalayer graphene has been reported[50].

## Methods

### Layer number determination

The layer number of graphene was determined through reflection contrast spectroscopy[51]. We mechanically exfoliated multilayer graphene from bulk crystals (NGS Naturgraphit) onto standard $SiO_2$/Si substrates, with the oxide layer thickness of 285 nm. Extended Data Fig. 1 shows an optical image of a 7L graphene flake with multiple steps at its edge. The optical contrast of various layers relative to the adjacent substrate near the edge is shown in Extended Data Fig. 1b. This optical contrast follows the Beer-Lambert law, exhibiting a linear increase relative to the layer number, as shown in Extended Data Fig. 1c. We find this thickness-dependent contrast is consistent across different flakes, allowing it to be utilized for identifying multilayer graphene even in the absence of stepped edges.

### Device fabrications

We fabricated high-quality rhombohedral 7L graphene using h-BN encapsulated structures with the assistant of the dry transfer method. The stacking order domains within the multilayer graphene were identified by Raman spectroscopy (WITec alpha300). During the dry transfer process, ABC stacking domains often shrink or even entirely convert to ABA stacking. To enhance the success rate, we isolated ABC domains from ABA domains by cutting the flake with a tungsten tip manipulated under a microscope. We found the cutting process did not significantly alter the domain distribution (see Extended Data Fig. 2). The entirely isolated ABC flake can survive after encapsulated by h-BN (see Extended Data Fig. 3b). Subsequently, we picked top h-BN and ABC graphene in sequence using a PDMS-supported PC film. The h-BN/graphene heterostructure was then released onto a bottom h-BN exfoliated on a 285 nm $SiO_2$/Si substrate in advance, forming the final stack.

To fabricate moiré superlattice devices, we intentionally align the crystallographic axes of graphene and h-BN by utilizing their straight edges. Typically, exfoliated large flakes of both graphene and h-BN, being hexagonal lattices, exhibit straight edges along their easy cleavage plane (either zig-zag or armchair). Extended Data Fig. 3a shows an optical image of the final stack for device D2. Notably, one of the natural cleavage edges of graphene is oriented perpendicular to the two straight edges of both the top and bottom h-BN, indicating that this stack is likely to be doubly aligned. Given the indistinguishable zig-zag and armchair edges, the aligned angle between graphene and h-BN can be around either 0° or 30°, which can be easily distinguished from transport data.

To verify the alive ABC domains in the final stack, we further characterized it by Raman spectroscopy as shown in Extended Data Fig. 3b. Pure ABC domains were carefully selected for the device design, with particular attention to regions devoid of bubbles, as determined through atomic force microscopy.

For the electrical contacts, we patterned the electrodes by e-beam lithography and selectively etched the top h-BN using $CHF_3/O_2$ plasma by controlling the etch duration. With this procedure, the multilayer graphene was exposed for metal deposition, thus forming 2D surface contacts. The electrodes and metallic top gates were fabricated by standard e-beam lithography and e-beam evaporation. The device was finally shaped into Hall bar geometry through drying etching with

$CHF_3/O_2$ plasma.

**Electronic transport measurements**

Low temperature transport measurements were performed in a dilution fridge (Oxford Triton) with a base temperature down to 50 mK. To minimize electronic temperature effects, all the wires were filtered by RC and RF filters (available from QDevil) at the mixing chamber. Standard low-frequency AC measurement was used to simultaneously obtain longitudinal and Hall resistances of the Hall bar device through lock-in amplifiers (SR830) operating at a frequency of 17.77 Hz. To measure the fragile ferromagnetic states, the AC current was limited within 5 nA. For other measurements, the current was increased to 100 nA to enhance signal quality. Gate voltages were applied using Keithley 2450 or 2614B.

The dual-gate structure of our devices provides independent control over both the total carrier density $n = n_t + n_n = \frac{C_t \Delta V_t}{e} + \frac{C_b \Delta V_b}{e}$ and the displacement field $D = \frac{C_t \Delta V_t - C_b \Delta V_b}{2\varepsilon_0}$, where $\Delta V_t = V_t - V_t^0$ ($\Delta V_b = V_b - V_b^0$) is the effective top (bottom) gate voltage, $V_t$ ($V_b$) the applied top (bottom) gate, $n_t$ ($n_b$) the top (bottom) gate induced carrier density, $V_t^0$ ($V_b^0$) the offset voltage, $C_t$ ($C_b$) the top (bottom) gate normalized capacitance measured from Hall effect at normal states, $e$ the elementary charge, and $\varepsilon_0$ the vacuum permittivity.

The measured Hall resistance $R_{xy}$ inevitably contains signals from longitudinal resistance $R_{xx}$ due to the unperfect geometry. To remove the components of $R_{xx}$ from $R_{xy}$, we used the standard procedure to anti-symmetrize Hall resistance ($\rho_{xy}$) by $\rho_{xy}(B, \leftarrow) = [R_{xy}(B, \leftarrow) - R_{xy}(-B, \rightarrow)]/2$ and $\rho_{xy}(B, \rightarrow) = [R_{xy}(B, \rightarrow) - R_{xy}(-B, \leftarrow)]/2$, where $\leftarrow$ and $\rightarrow$ represent the swept magnetic field from positive to negative and from negative to positive, respectively. The residual resistance is defined as $\Delta\rho_{xy}^{AH} = [\rho_{xy}(B = 0, \leftarrow) - \rho_{xy}(B = 0, \rightarrow)]/2$.

**Band structure calculation**

Rhombohedral 7L graphene with moiré superlattice has Hamiltonian

$$H_{tot} = H_7 + V_{mo},$$

where $H_7$ is the effective tight-binding Hamiltonian of the intrinsic rhombohedral 7L graphene, and the effective intralayer moiré potential $V_{mo}$ is only applied to the graphene layer contacting with h-BN layer[52].

Using the Slonczewski-Weiss-McClure (SWMC) tight-binding lattice model, the Hamiltonian can be written as

$$H_7 = \begin{pmatrix} D_1 & V & W & 0 & 0 \\ V^\dagger & D_2 & V & \ddots & 0 \\ W^\dagger & V^\dagger & \ddots & \ddots & W \\ 0 & \ddots & \ddots & D_6 & V \\ 0 & 0 & W^\dagger & V^\dagger & D_7 \end{pmatrix} \dots\dots\dots\dots (1)$$

where the 2×2 blocks are

$$D_i = \begin{pmatrix} u_{Ai} + \delta_i & v_0\pi^\dagger \\ v_0\pi & u_{Bi} + \delta_i \end{pmatrix},$$

$$V = \begin{pmatrix} -v_4\pi^\dagger & -v_3\pi \\ t_1 & -v_4\pi^\dagger \end{pmatrix},$$

$$W = \begin{pmatrix} 0 & t_2 \\ 0 & 0 \end{pmatrix}.$$

Here $v_i = \sqrt{3}a_G t_i/(2\hbar)$ $(i = 0,3,4)$ and the subscripts $Ai, Bi$ reprпоявent two sublattices in $i$th layer. The term $\pi = \hbar(\nu k_x + ik_y)$ is defined by the valley index $\nu = \pm 1$ using the wave vector $\vec{k} = (k_x, k_y)$ measured from Dirac points $K_\nu = \left(\frac{4\pi\nu}{3a_G}, 0\right)$. The diagonal site potentials are

$$u_{A1} = u_{B7} = 0\ eV$$
$$u_{A7} = u_{B1} = 0.0122\ eV$$
$$u_{Ai} = u_{Bi} = -0.0164\ eV\ (1 < i < 7),$$

And $\delta_i = (4 - i)\Delta$ (for $1 \le i \le 7$) introduces the interlayer potential difference ($\Delta$) between contiguous layers through a perpendicular external electric field. The effective tight-binding parameters are $t_0 = 3.1$ eV, $t_1 = 0.3561$ eV, $t_2 = -0.0083$ eV, $t_3 = 0.293$ eV, $t_4 = 0.144$ eV, which represent hopping terms between different sites.

The character of moiré superlattice is captured by adding $V_{mo}$ acting on both (or one of) the top and bottom layers of rhombohedral 7L graphene[53]. The reciprocal lattice vectors of graphene ( $\vec{g}^G$), h-BN ($\vec{g}^{BN}$) and moiré superlattice ($\vec{G}$) are respectively given by

$$\vec{g}_m^G = \hat{R}_{\frac{\pi(m-1)}{3}}\left(0, \frac{4\pi}{\sqrt{3}a_G}\right)^{\mathrm{T}}$$

$$\vec{g}_m^{BN} = \hat{R}_{\frac{\pi(m-1)}{3}}\left(0, \frac{4\pi}{\sqrt{3}a_{BN}}\right)^{\mathrm{T}}$$

$$\vec{G}_m = \hat{R}_\theta \vec{g}_m^{BN} - \vec{g}_m^G, \qquad m \in \{1,2,\ldots,6\},$$

where $a_{BN} = 0.250$ nm is the lattice constant of h-BN. The $\theta$ represents the relative twist angle between the graphene and h-BN layers. $\hat{R}_\varphi = \begin{pmatrix} \cos\varphi & -\sin\varphi \\ \sin\varphi & \cos\varphi \end{pmatrix}$ rotates a vector by angle $\varphi$. We use $\xi = \pm 1$ to distinguish the two possible alignments between graphene and h-BN. It represents the perturbation of the low-energy A (bottom) or B (top) sites in graphene by h-BN in two different ways, giving rise to different band structures[52].

Then, we can express the $V_{mo}$ operator as a matrix in $\boldsymbol{k}$-space with basis (*A1, B1*) or (*A7, B7*),

$$< \vec{k} + \xi\vec{G}_m\, |V_{mo}|\vec{k} >$$
$$= V_{AA}^\xi\left(\frac{I + \xi\sigma_z}{2}\right) + V_{BB}^\xi\left(\frac{I - \xi\sigma_z}{2}\right) + \frac{\left[V_{BA}^\xi\delta_{\nu,1} + V_{AB}^\xi\delta_{\nu,-1}\right](\sigma_x, \xi\sigma_y)M\vec{G}_m}{|\vec{G}_1|} \quad \ldots(2)$$

where $\sigma_{x,y,z}$ are the Pauli matrices,

$$V_{AA}^\xi = C_{AA}\, e^{(-1)^{m+1}\xi\phi_{AA}i}$$
$$V_{BB}^\xi = C_{BB}\, e^{(-1)^{m+1}\xi\phi_{BB}i}$$
$$V_{BA}^\xi = \left(V_{AB}^\xi\right)^* = -\xi C_{AB} e^{(-1)^{m+1}\xi\left(\phi_{AB} - \frac{\pi}{6}\right)i}$$

$$M = \frac{1}{\sqrt{\alpha^2 - 2\alpha\cos\theta + 1}} \begin{pmatrix} -\alpha\sin\theta & -\alpha\cos\theta + 1 \\ \alpha\cos\theta - 1 & -\alpha\sin\theta \end{pmatrix}$$

with $C_{AA} = -14.88$ meV, $C_{BB} = 12.09$ meV, $C_{AB} = 11.34$ meV, $\phi_{AA} = 50.19°$, $\phi_{BB} = -46.64°$, $\phi_{AB} = 19.6°$, $\alpha = a_G/a_{BN}$.

The band calculation of the intrinsic rhombohedral 7L graphene is very similar to that of bilayer graphene. As for the moiré superlattice, the original band are reconstructed into a small moiré Brillouin zone (MBZ), which is a hexagon with its center ($\tilde{\Gamma}$ point) and one corner ($\tilde{K}$ point) located at adjacent corners of the BZs of graphene and h-BN, respectively. To calculate the band structure in the MBZ, for each momentum $\vec{k}$ in it, we build a large matrix $H(\vec{k})$ whose bases include states for $3n(n+1)+1$ momentum points $\vec{k}'$ which satisfy

$$\vec{k}' = \vec{k} + \sum_{m=1}^{6} c_m \vec{G}_m, \qquad c_m \in \mathbb{N}, \quad \sum_{m=1}^{6} c_m \le n.$$

Here $n$ is the truncation length. The diagonal blocks in $H(\vec{k})$ are copies of Hamiltonian in Eq. (1) with different momentum $\vec{k}'$. And the non-diagonal blocks is where the intralayer moiré potential $V_{mo}$ performs, following Eq. (2). Selecting $n = 2$ and diagonalizing $H(\vec{k})$ give the precise enough energy dispersion.

**Raman spectroscopy**

To elucidate the impact of moiré potential on the surface and bulk states of 7L graphene, we conducted a comparative analysis of Raman spectra. We examined both aligned and non-aligned 7L graphene, encompassing ABA and ABC stacking domains.

It's well known that in the case of monolayer and bilayer graphene, the moiré superlattice formed between graphene and h-BN can induce a periodic strain distribution, resulting in a noticeable broadening of the 2D peak[54,55]. For instance, in monolayer graphene, when singly aligned with h-BN, the full width at half-maximum (FWHM) of the 2D peak exhibits a value of approximately 20 $cm^{-1}$ larger than non-aligned counterpart. This broadening becomes even more pronounced, reaching an increase of about 40 $cm^{-1}$ for doubly aligned monolayer graphene in comparison to the non-aligned one[56].

However, in the case of ABC graphite, we observe no distinguished broadening of the 2D peak between aligned and non-aligned configurations. As shown in Extended Data Fig. 3c, the FWHM exhibits similar value of ~ 70 $cm^{-1}$. This feature is consistent with a recent report on ABA-stacked graphite[46], which also demonstrates a very short penetration depth for moiré reconstruction.

**Twist angle determination**

To accurately determine $\theta$, we employed Brown-Zak oscillations occurring at a moiré superlattice under high magnetic fields ($B$)[38,40]. In Fig. 2c, we observed remarkable quantum oscillations with periodicities independent of $n$ with a fixed $D = 0$. Further analysis reveals that the minima in $R_{xx}$ exhibit two sets of oscillatory behavior periodic in $1/B$, characteristic of Brown-Zak oscillations[42]. In systems with superlattices under a magnetic field, the electronic spectra can develop into fractal spectra known as Hofstadter butterflies, resulting in a series of minima $R_{xx}$ at $B = \phi_0/qA$, where $q$ is a integer, $\phi_0$ magnetic flux quantum, and $A = \sqrt{3}\lambda^2/2$ the unit-cell area of the superlattice. The low resistance observed in Brown-Zak oscillations stems from the repetitive formation of magnetic Bloch states at magnetic field following the sequence of $\frac{\phi}{\phi_0} = 1/q$, in which electrons recover delocalized wave functions and propagate along open trajectories instead of cyclotron trajectories. The two distinct sets of Brown-Zak oscillations in Fig. 2c, indicate that both the top and bottom h-BN are aligned with graphene in device D2[43,44]. By individually fitting these oscillations, we extracted two moiré wavelengths as $\lambda_1 = 11.1$ nm and $\lambda_2 = 11.0$ nm, corresponding to two twist angles $\theta_1 = 0.88°$ and $\theta_2 = 0.90°$ at the top and bottom interfaces.

The $\theta$ can also be calculated from resistance peaks at $n_1$ and $n_2$ corresponding to $\nu = 1$ and $\nu = 2$. Given the four-fold degeneracy (two for spin and two for valley) in graphene, four electrons per moiré cell are required for full filling of a moiré miniband ($\nu = 4$). The corresponding $n$ at $\nu = 4$ is $n_s = 4n_1 = 2n_2 \approx 8\theta^2/\sqrt{3}a_G^2$. The twist angles calculated from this method are 0.90° and 0.94°, approximately consistent with those extracted from Brown-Zak oscillations.

**LLs at high field**

In ABA-stacked graphite thin film, at high *B*, when the system enters into the ultra-quantum regime, only the two lowest Landau bands (0 and 1) are across the Fermi energy. Within this regime, electrons form a set of standing waves along the c-axis, penetrating across the entire bulk owing to the one-dimensional Landau bands because of a finite thickness of graphite films. These states are thus subject to the influence of both top and bottom gates, resulting in the diagonal features at center of $R_{xx}(n_t, n_b)$ map. Meanwhile, the horizontal and vertical features at the edge of $R_{xx}(n_t, n_b)$ map are attributed to quantized states at the graphite surfaces, coexisting with the screening bulk states[45,46]. These features occur at both standard quantum Hall effect and Hofstadter's butterfly gaps, evidenced by recent observations that the moiré surface potential affects the entire bulk of graphite in the ulta-quantum regime[46,47].

In rhombohedral multilayer graphene, the situation is quite different. At moderate *B*, The Landau diagrams plotted in $R_{xx}(n_t, n_b)$ (as shown in Fig. 3) exhibit a series of horizontal and vertical features at small *D*, whereas these features transform into diagonal patterns at high *D*. These characteristics reveal the surface states and 3D features inherent to rhombohedral 7L graphene. At low temperature, the electronic transport properties of rhombohedral multilayer graphene are dominated by its surface states, with the conductivity through the 3D bulk band being effectively suppressed[6]. Specifically, at small *D*, there is a finite band overlap between the conduction and valence bands, leading to the presence of two metallic surface states (the correlation gap at *n*=0, *D*=0 is smeared out by the strong magnetic field). In this scenario, the gate-voltage-induced surface charge accumulation on one of the two surfaces exerts a strong screening effect, diminishing its gating effect on the other surface. Consequently, the two surface states become electronically decoupled. However, at high *D*, the inversion symmetry breaking overcomes the band overlap, effectively polarizing the surface states. When fixing *D* and tuning *n*, only one of the two surfaces become conductive, and this conductivity can be effectively tuned by both gates. As a result, we observe the emergence of diagonal LLs at high *D*, as shown in Fig. 3.

**$N\pi$ Berry phase**

In rhombohedral $N$-layer graphene, low-energy electrons are localized on the top and bottom layers. A simplified two-band model approximately describes the low-energy bands with a dispersion relation of $E \sim \pm p^N$, exhibiting a large DOS at $n \to 0$ and a $\pm N\pi$ Berry phase around K/K' valley. The valley and layer are strongly coupled such that K (K') valley associated with $N\pi$ $(-N\pi)$ Berry phase is localized on the top (bottom) layer. Under the influence of $B$, the low-energy bands in rhombohedral graphene develop into a series of LLs. The zeroth LL has $N$-fold orbital degeneracy, two-fold valley degeneracy, and two-fold spin degeneracy. This LL evolves with $B$ as $E^{\pm} \propto \pm B^{N/2}$. It's worth noting that the valley degeneracy in the zeroth LL is equivalent to layer degeneracy. In the case of rhombohedral 7L graphene, the zeroth LL has a total degeneracy of 28, corresponding to the filling factor from -14 to +14. The application of an external $D$ breaks the inversion symmetry and lifts the valley degeneracy. Consequently, the zeroth LL is split as -7 to +7. This layer-number dependent orbital degeneracy serves as an indicator of the layer number in rhombohedral multilayer graphene devices. Additionally, it also provides evidence for the existence of $N\pi$ Berry phase in rhombohedral $N$-layer graphene.

Experimentally, high-order hopping terms distort the simple power-law low-energy bands in terms of trigonal warping and electron-hole asymmetry, which break the orbital degeneracy of the zeroth LL and induce a series of LLs crossings at the hole side.

In Extended Data Fig. 6, we present the observed manifestation of a $7\pi$ Berry phase in rhombohedral 7L graphene, in the absence of a moiré superlattice. Specifically, in the absence of external displacement filed ($D = 0$ V nm$^{-1}$), pronounced $\nu = -14$, corresponding to the gap between zeroth LL and first LL, is marked. Within the zeroth LL, we observed both orbital splitting and spin splitting at high $B$. Within the region of $n$ between $-3.5 \times 10^{12}$ cm$^{-2}$ and $-5.5 \times 10^{12}$ cm$^{-2}$, the zeroth LL overlaps with the valence-band LLs, resulting in intricate quantum oscillations. At sufficiently high hole density and large $B$, the simple four-fold degeneracy characteristic of valence-band LLs is recovered. The introduction of a finite $D$ results in the lifting of valley degeneracy due to the breaking of inversion symmetry. As a result, pronounced $\nu = -7$ emerges, confirming the presence of a $7\pi$ Berry phase in rhombohedral 7L graphene.

In general, the quantum Hall states observed in rhombohedral 7L graphene resemble those reported in rhombohedral 9L or thicker graphene previously reported[6]. It is essential to emphasize that rhombohedral 7L graphene effectively represents a transition from 2D characteristics to 3D attributes in rhombohedral graphene systems.

**Spontaneous symmetry breaking in non-aligned rhombohedral 7L graphene**

In intrinsic rhombohedral 7L graphene without moiré, the surface flat band near Fermi surface favors interaction-driven symmetry breaking. In the non-interacting regime, this band has a four-fold degeneracy due to the presence of spin and valley symmetries. This can be revealed from the period of Shubnikov–de Haas (SdH) oscillations at relatively low $B$, calculated as $\Delta\nu = \frac{\Delta n h}{eB}$. As shown in Extended Data Fig. 7b, at high $n$ we observed $\Delta\nu = 4$, consistent with expectations from the single-particle picture. The application of $D$ can further flatten the surface band, dramatically increasing DOS near vHSs. When the Stoner criterion $UD_F > 1$ is satisfied at specific $n$ and $D$, spontaneous spin-valley flavor polarization occurs. In this situation, the initial four-fold degeneracy is reduced to two-fold in the case of a half-metal state (spin or valley polarized) or fully lifted in the case of a quarter-metal state (spin and valley polarized). This degeneracy lifting can be observed both in the $n$-$D$ mapping at a fixed $B$ (Extended Data Fig. 7b) and in quantum oscillations as a function of $B$ (Extended Data Fig. 7c). When in a half-metal state, we observed AHE, providing evidence of valley polarization occurring as $B$ approaches zero. This valley polarization gives rise to a nonzero Berry curvature, leading to an intrinsic AHE. It's worth noting that the valley-polarized half-metal state observed in this study is distinct from those in thinner rhombohedral graphene systems[7,32,33], where AHE was only observed in quarter-metal region. Our observations in 7L graphene offer valuable insights into the evolution of Stoner instability with increasing layer number. Notably, the pronounced screening effect between two surface states observed in the low $D$ region (see Fig. 3 and Extended Data Fig. 7b), absent in previously reported thinner rhombohedral graphene, strongly indicates that 7L graphene represents a the crossover from 2D to 3D systems.

Though the half-metal state with AHE emerges in the intrinsic rhombohedral 7L graphene without a moiré superlattice, it's noted that this state appeared at a very narrow region. The introduction of moiré superlattice in rhombohedral 7L graphene can further flatten the surface band and favor the spontaneous symmetry breaking. This, in turn, facilitates the emergence of surface ferromagnetism across a significantly large region, as shown in Fig. 4.

**Acknowledgements:**
This work was funded by National Natural Science Foundation of China (Grant No. 12274354), the R&D Program of Zhejiang province (2022SDXHDX0005), and Westlake Education Foundation at Westlake University. We thank Chao Zhang from the Instrumentation and Service Center for Physical Sciences (ISCPS) at Westlake University for technical support in data acquisition. We also thank Westlake Center for Micro/Nano Fabrication and the Instrumentation and Service Centers for Molecular Science for facility support. K.W. and T.T. acknowledge support from the JSPS KAKENHI (Grant Numbers 21H05233 and 23H02052) and World Premier International Research Center Initiative (WPI), MEXT, Japan.

**Author Contributions**
S.X. conceived the idea and supervised the project. W.Zhou fabricated the devices. J.D. built up the measurement system. W.Zhou performed the transport measurement with the assistance of J.D. and L.Z.. J.H. and W.Zhu performed the band structure calculations. K.W. and T.T. grew h-BN crystals. S.X. wrote the paper with the input from W.Zhou, W.Zhu, and J.H.. All authors contributed to the discussions.

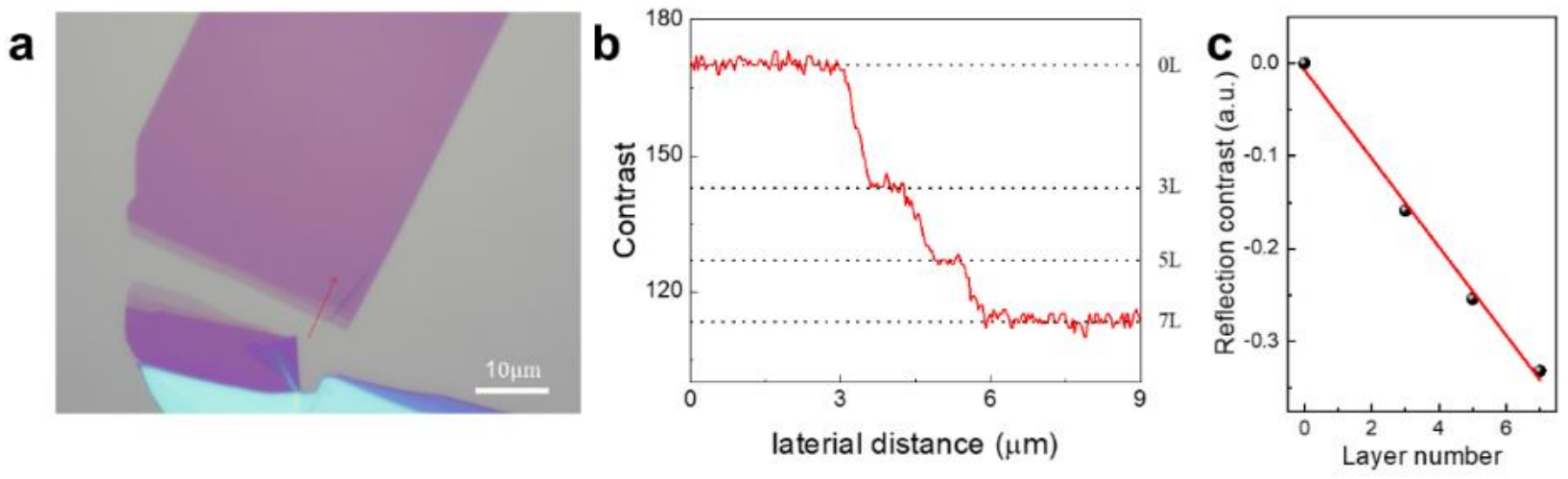

**Extended Data Fig. 1 | Identification of layer number of multilayer graphene**. **a**, Optical image of a typical 7L graphene. **b**, Cross-sectional profile of optical contrast along the red line in (**a**). The step-like edge helps us identify the layer number accurately. **c**, layer-dependent reflection contrast. The red line is a fit to the date using the Beer-Lambert law. The layer-number-dependent optical contrast facilitates the determination of layer number quickly and reliably, compared with other techniques such as Raman spectroscopy.

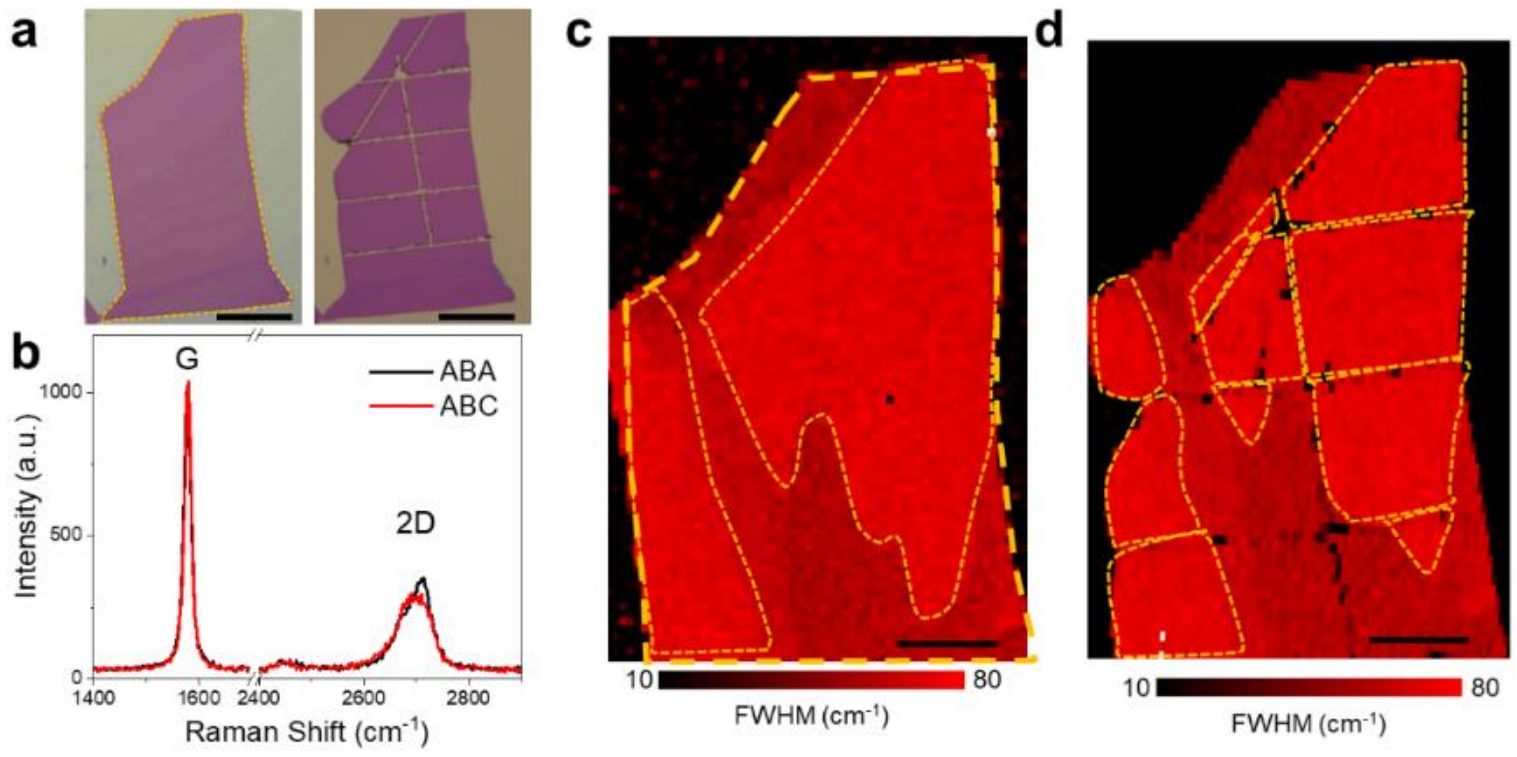

**Extended Data Fig. 2 | Raman characterization before and after cutting process**. **a**, Optical images of multilayer graphene before and after cut by a tungsten tip. **b**, Comparison of Raman spectra acquired in ABA and ABC domains. The stacking order can be identified through the shape of 2D peak and the position of G peak. **c**, **d**, Raman mapping of the full width at half maximum (FWHM) of 2D peak for the flake before (**c**) and after (**d**) cut. The cutting process does not significantly change the domain distribution. Therefore, it can be used to stabilize ABC domain during the transfer process. The scale bars in (**a**) are both 20 μm. The scale bars in (**c**) and (**d**) are 10 μm.

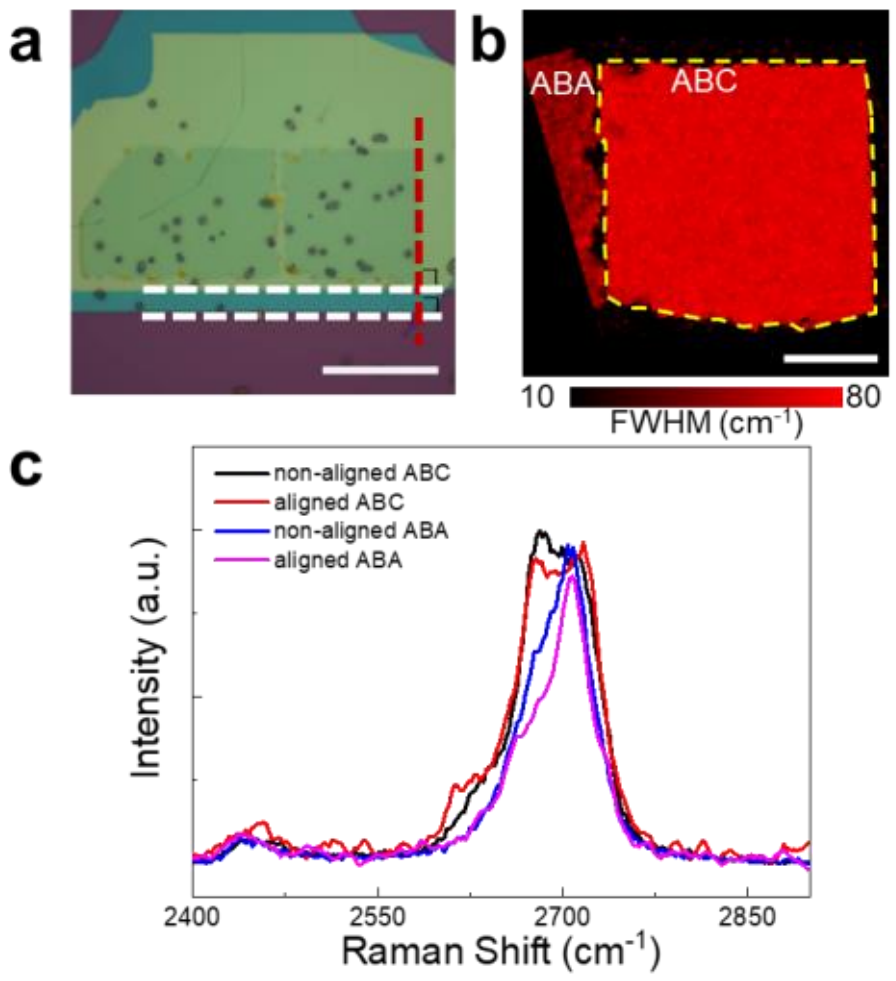

**Extended Data Fig. 3 | Raman characterization of h-BN encapsulated rhombohedral 7L graphene**. **a**, Optical image of the 7L graphene encapsulated by h-BN. The dashed lines mark the alignment between graphene, top and bottom h-BN. The straight edge of graphene is perpendicular to both top and bottom h-BN, indicating the stack is doubly aligned. This stack is the one for device D2 in the main text. The scale bar is 25 μm. **b**, Raman map of the full width half maximum (FWHM) of 2D band peak for the sample in (**a**). The scale bar is 10 μm. **c**, Normalized Raman spectra around 2D band peak for aligned and non-aligned ABA- and ABC-stacked 7L graphene. All the samples are encapsulated by h-BN.

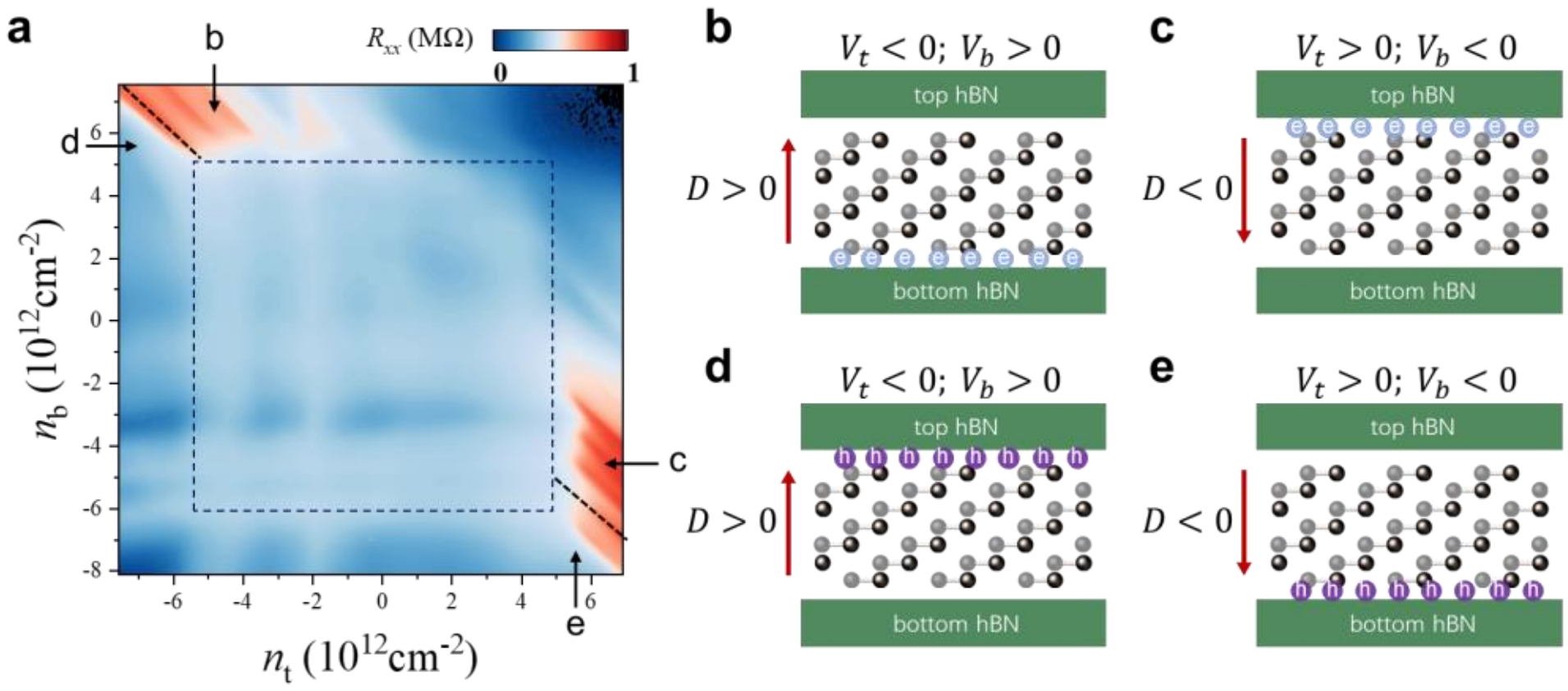

**Extended Data Fig. 4 | Surface states and screening effect in aligned rhombohedral 7L graphene**. **a**, Longitudinal resistance $R_{xx}$ as a function of carrier density $n_t$ and $n_b$ induced by top and bottom gate, respectively, measured at $T$=50 mK and $B$=0 T. The data are the same as that in Fig. 2b, but plotted as $n_t = \frac{C_t \Delta V_t}{e}$ and $n_b = \frac{C_b \Delta V_b}{e}$. A series of horizontal and vertical lines were observed inside the region marked by the dashed box, which indicates that the two surface states are electronically decoupled and effectively screened out by each other. Under large displacement field ($|D| > |D_c|$), layer-polarized surface states dominate, namely, only one of the two surfaces contributes to the conduction and the other one is fully depleted. The four kinds of layer-polarized surface states marked by the arrows are schematically shown in **b-e**, respectively. In these states, both gates can effectively tune the individual surface state.

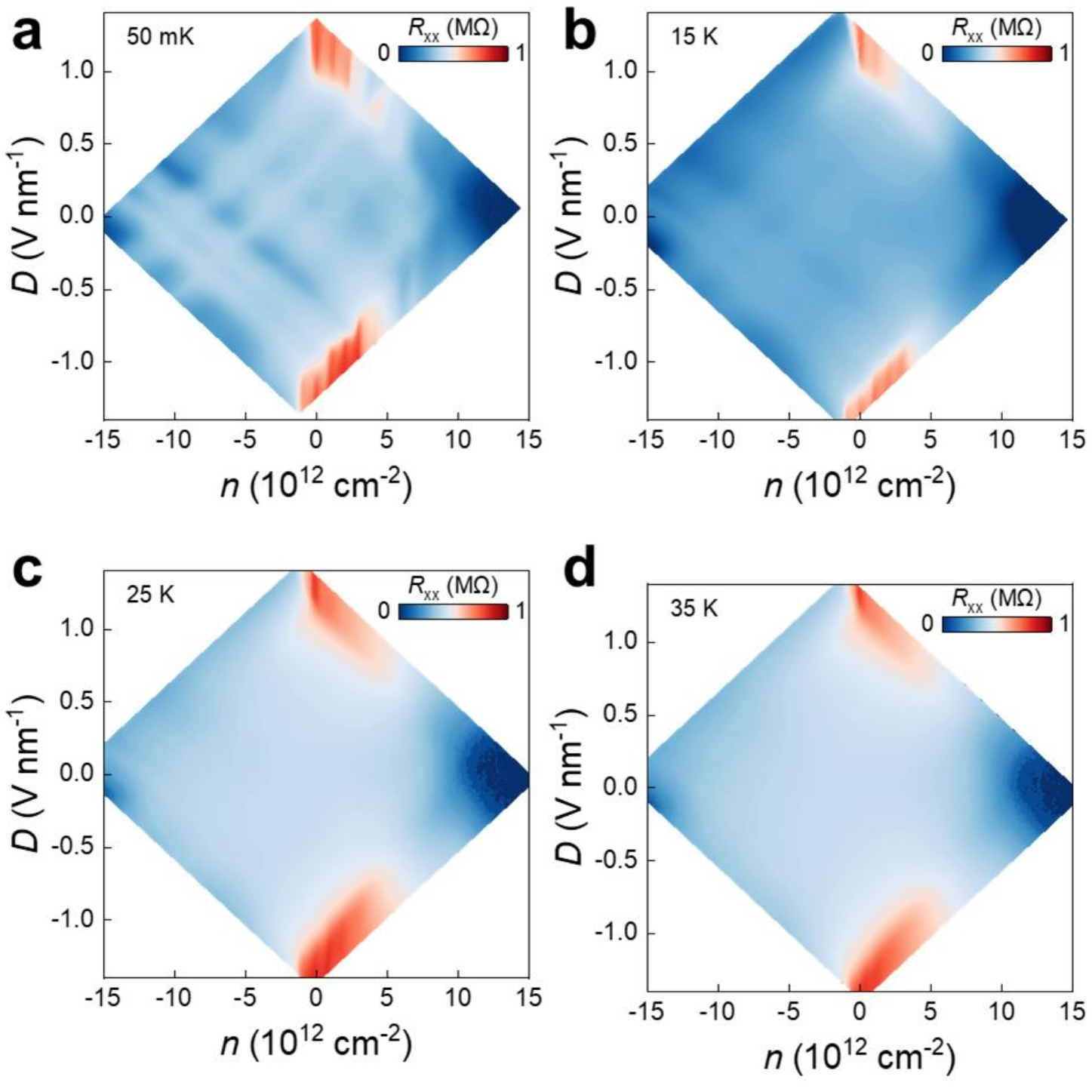

**Extended Data Fig. 5 | Temperature dependence of $R_{xx}(n,D)$ mapping for the aligned rhombohedral 7L graphene**. Color maps of longitudinal resistance $R_{xx}$ as a function of carrier density *n* and displacement field *D* measured at *T*=50 mK (**a**), 15 K (**b**), 25 K (**c**), 35 K (**d**). The data were measured in Device D2 at $B = 0$ T.

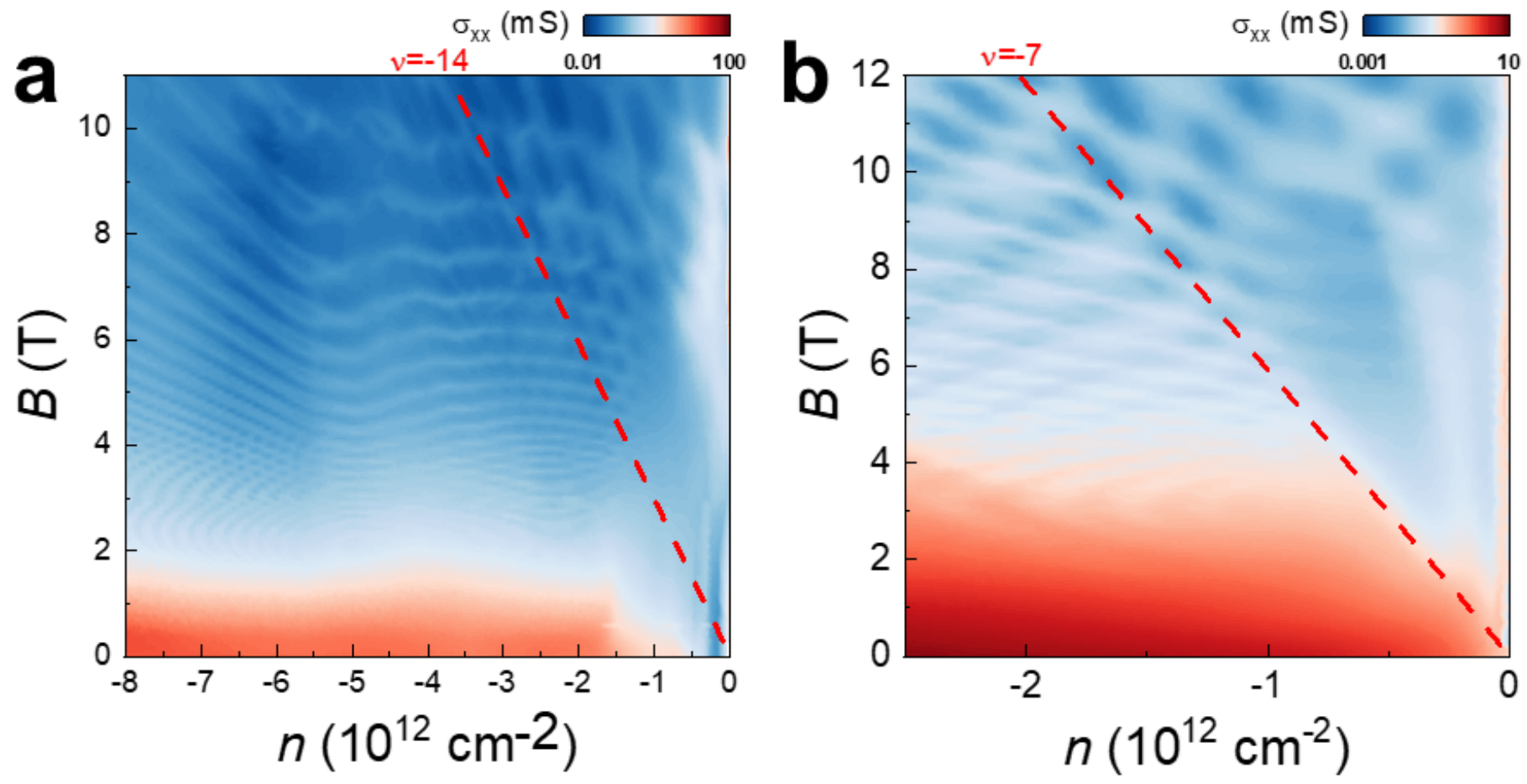

**Extended Data Fig. 6 | Landan fan diagrams of non-aligned rhombohedral 7L graphene**. **a**, **b**, Longitudinal conductivity $\sigma_{xx}(n,B)$ mapping as a function of *n* and *B* at fixed *D*=0 V nm$^{-1}$ (**a**) and *D*=-0.55 V nm$^{-1}$ (**b**). At *D*=0 V nm$^{-1}$, pronounced $\nu = -14$ corresponding to the gap between zeroth LL and first LL in the valence band was observed. Under non-zero *D*, valley degeneracy is lifted due to the inversion symmetry breaking, leading to the appearance of pronounced $\nu = -7$. Both (**a**) and (**b**) demonstrate the existence of $7\pi$ Berry phase in intrinsic rhombohedral 7L graphene. The data were measured in Device D3.

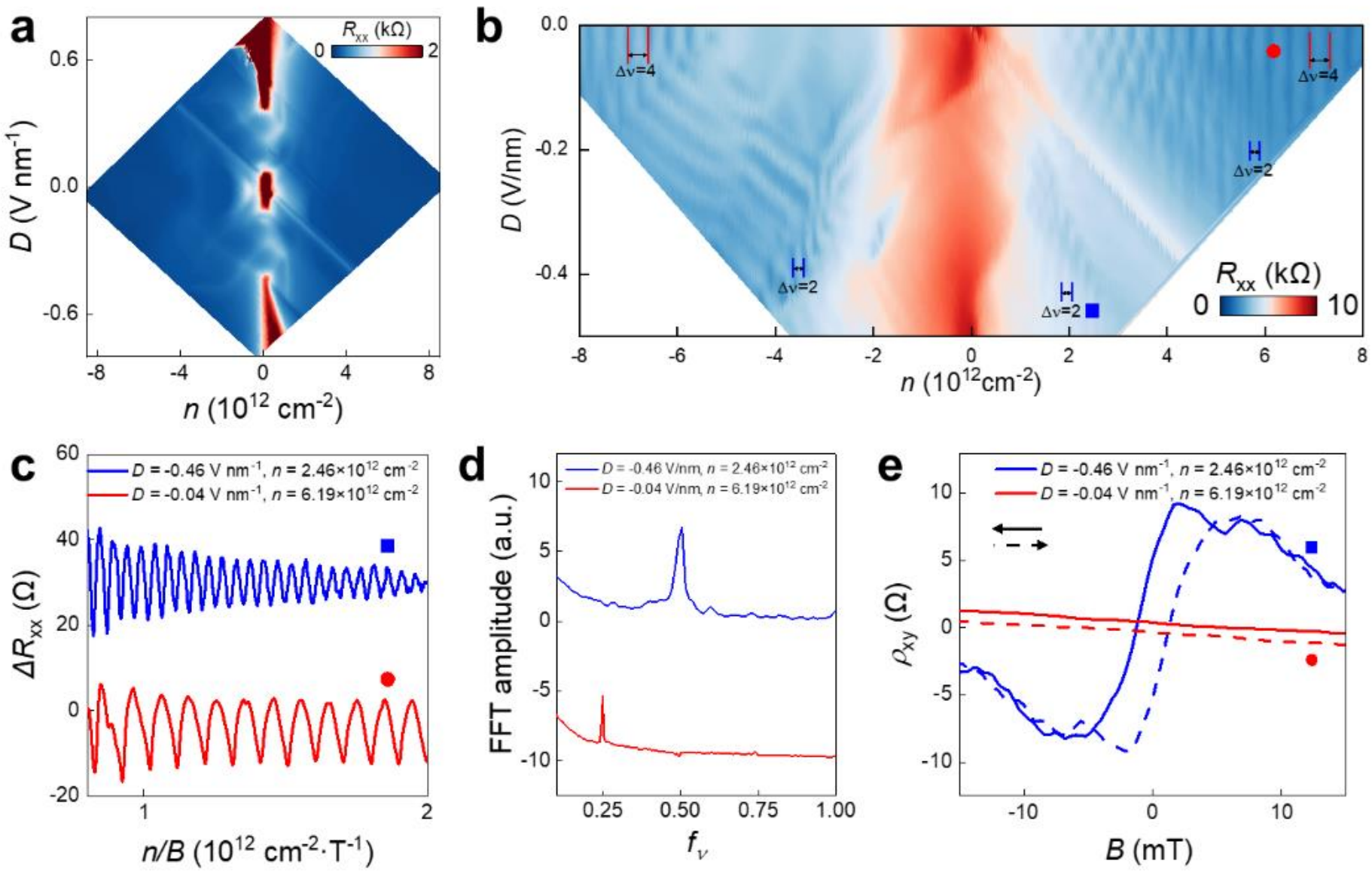

**Extended Data Fig. 7 | Spontaneous symmetry breaking in non-aligned rhombohedral 7L graphene. a,** Color maps of longitudinal resistance $R_{xx}$ as a function of carrier density $n$ and displacement field $D$ measured at $B$ =0 T for the devices without moiré superlattice (Device 3). **b**, $R_{xx}(n, D)$ mapping at $B$ = 4 T. Quantum oscillations with different degeneracies were observed. At high $n$, typical degeneracy of 4 (2 spins × 2 valleys) in graphene is marked with red lines. At low $n$, half metal with degeneracy of 2 was observed, which is marked with blue lines. **c**, SdH oscillations at normal and half metal states. The data were taken at the position marked with circle (red) and square (blue) labels in (**b**). **d**, The corresponding fast Fourier transform (FFT) of the SdH oscillations in (**c**). The data are plotted as a function of $f_\nu = f_B/(\phi_0 n)$, where $f_B$ is the oscillation frequency in the unit of tesla, $\phi_0 = h/e$ is the magnetic flux quantum. **e**, Low-field anti-symmetrized Hall resistance $\rho_{xy}$ as a function of $B$ measured at normal and half metal states labeled as those in (**c**) and (**d**). AHE with hysteresis loops was observed in the half metal state, indicating it is spontaneous valley polarized state. All the data were taken at $T$ =50 mK in Device D3.

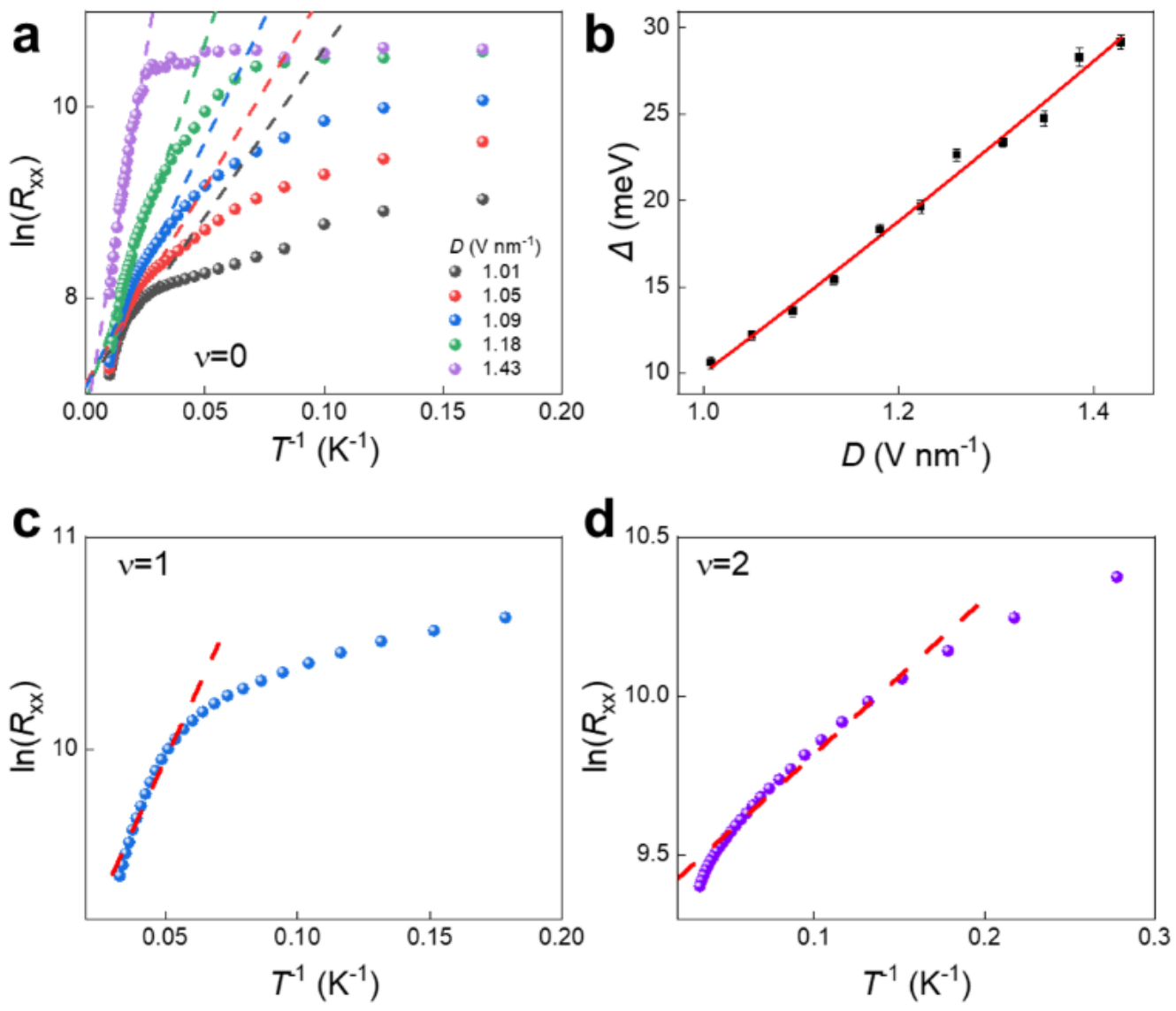

**Extended Data Fig. 8 | Arrhenius plot and extracted gap size of aligned rhombohedral 7L graphene.** **a**, $\ln(R_{xx})$ as a function of $T^{-1}$ under various displacement field at charge-neutrality points ($\nu = 0$). The dashed lines are the linear fits, which can be used to extract transport gap according to thermal activation equation $1/R_{xx} \propto e^{-\Delta/2k_BT}$. **b**, The measured gap size as a function of $D$ at charge-neutrality points. **c**, **d**, $\ln(R_{xx})$ as a function of $T^{-1}$ at quarter filling ($\nu = 1$) and half filling ($\nu = 2$) under $D$=1.1 V nm$^{-1}$. Linear fittings are obtained at low temperature regions. These analyses are based on the data in Device D2.

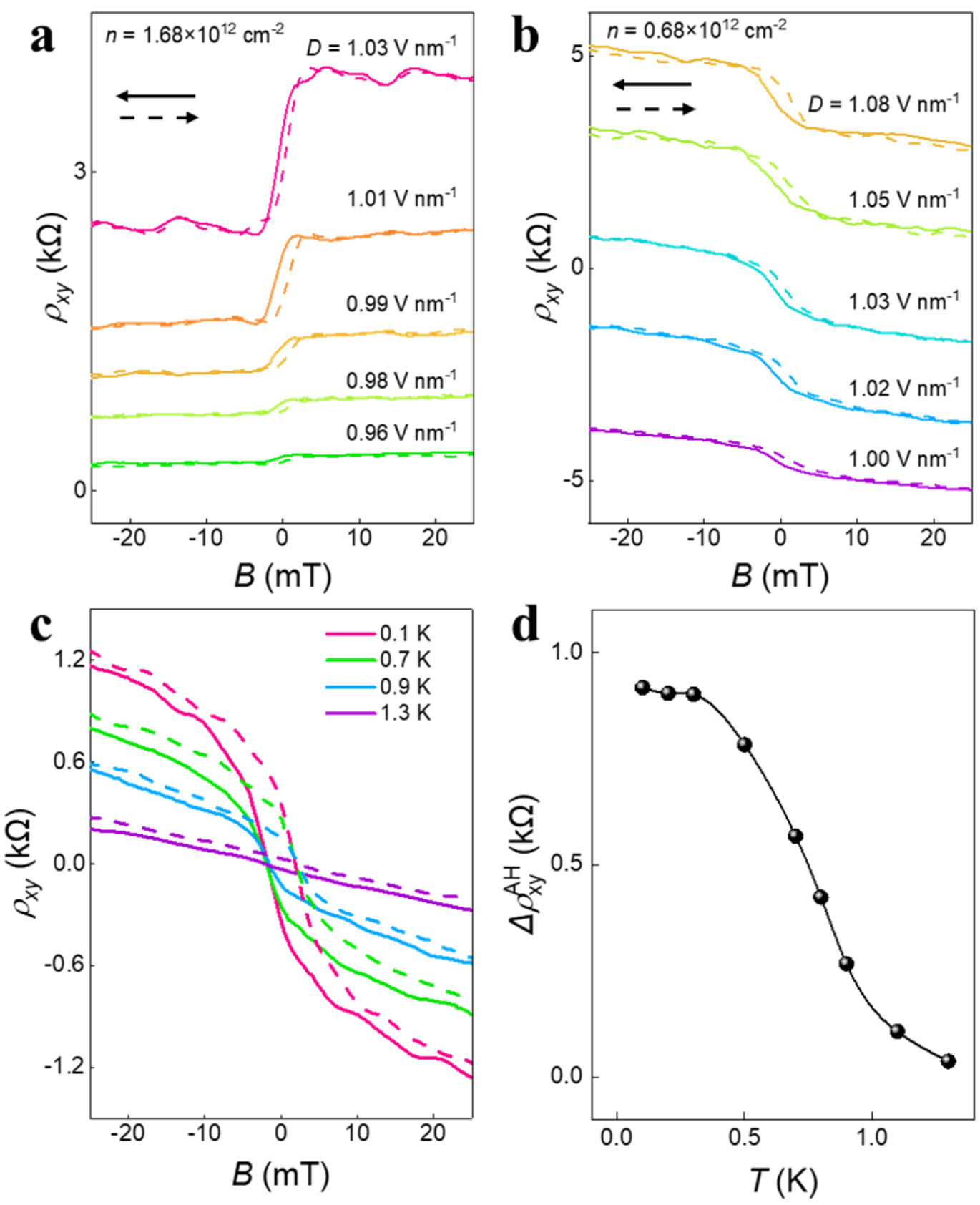

**Extended Data Fig. 9 | More data of AHE with hysteresis loops**. **a**, **b**, Anti-symmetrized Hall resistance $\rho_{xy}$ as a function of $B$ swept back and forth at low field regions at (**a**) a fixed $n = 1.68 \times 10^{12}$ cm$^{-2}$, varying $D$ from 1.03 V nm$^{-1}$ to 0.96 V nm$^{-1}$ and (**b**) a fixed $n = 0.68 \times 10^{12}$ cm$^{-2}$, varying $D$ from 1.08 V nm$^{-1}$ to 1.00 V nm$^{-1}$. The absolute values are manually offset for clarity. **c**, Temperature dependence of AHE. $\rho_{xy}$ as a function of $B$ swept back and forth, showing ferromagnetic hysteresis at different temperatures, with fixed $n = 0.67 \times 10^{12}$ cm$^{-2}$ and $D = 1.03$ V nm$^{-1}$. **d,** The corresponding residual Hall resistance $\Delta\rho_{xy}^{AH}$ as a function of temperature.

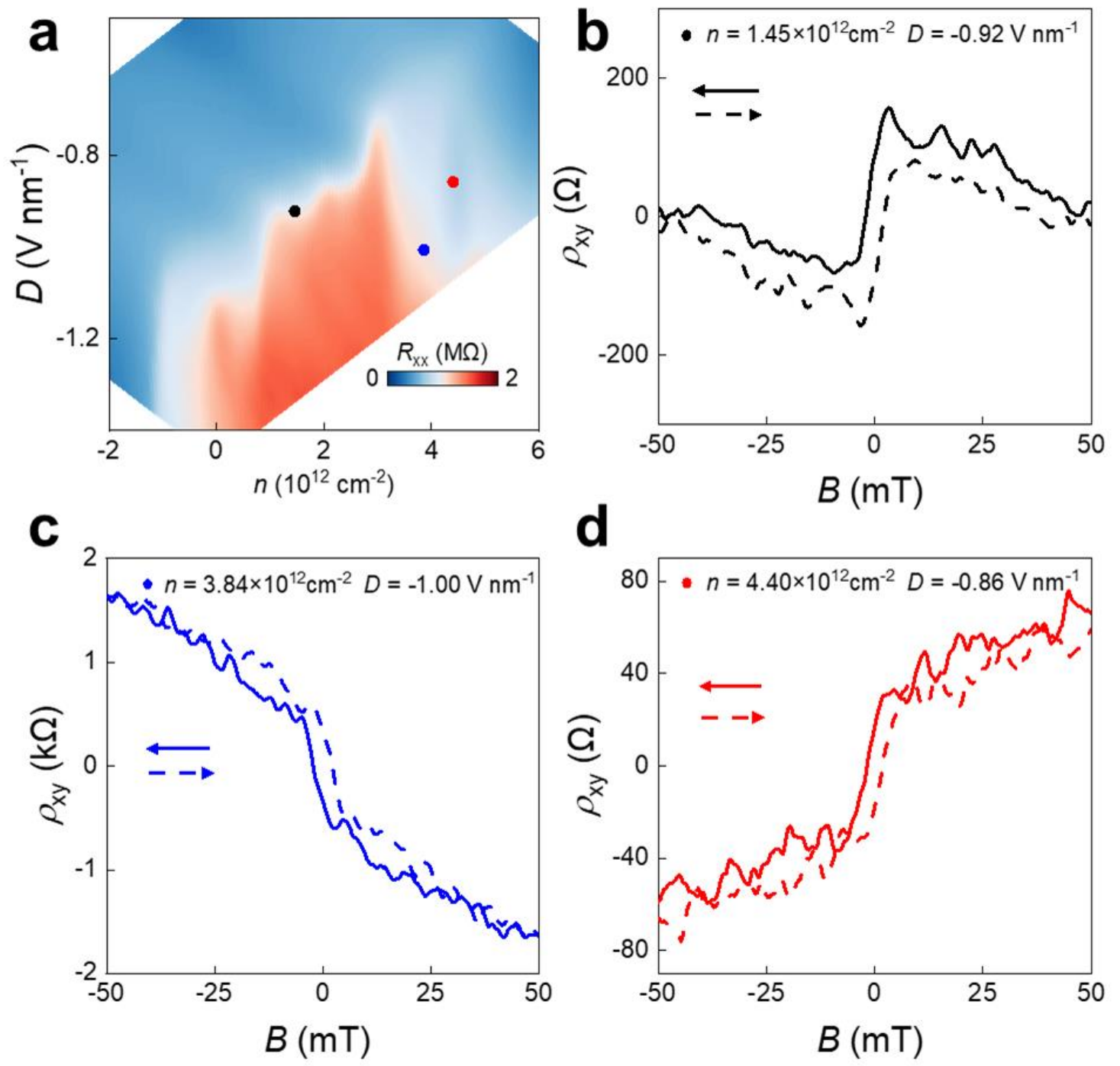

**Extended Data Fig. 10 | Ferromagnetic states at negative *D*.** **a**, Fine $R_{xx}(n, D)$ mapping at negative $D$ side. **b-d**, Anti-symmetrized Hall resistance $\rho_{xy}$ as a function of $B$ swept back and forth at three position marked with different colors in (**a**): $n = 1.45 \times 10^{12}$ cm$^{-2}$, $D = -0.92$ V nm$^{-1}$ (**b**); $n = 3.84 \times 10^{12}$ cm$^{-2}$, $D = -1.00$ V nm$^{-1}$ (**c**); $n = 4.40 \times 10^{12}$ cm$^{-2}$, $D = -0.86$ V nm$^{-1}$ (**d**). AHE with both nonlinear features and hysteresis loops was observed. The data were taken in Device D2.

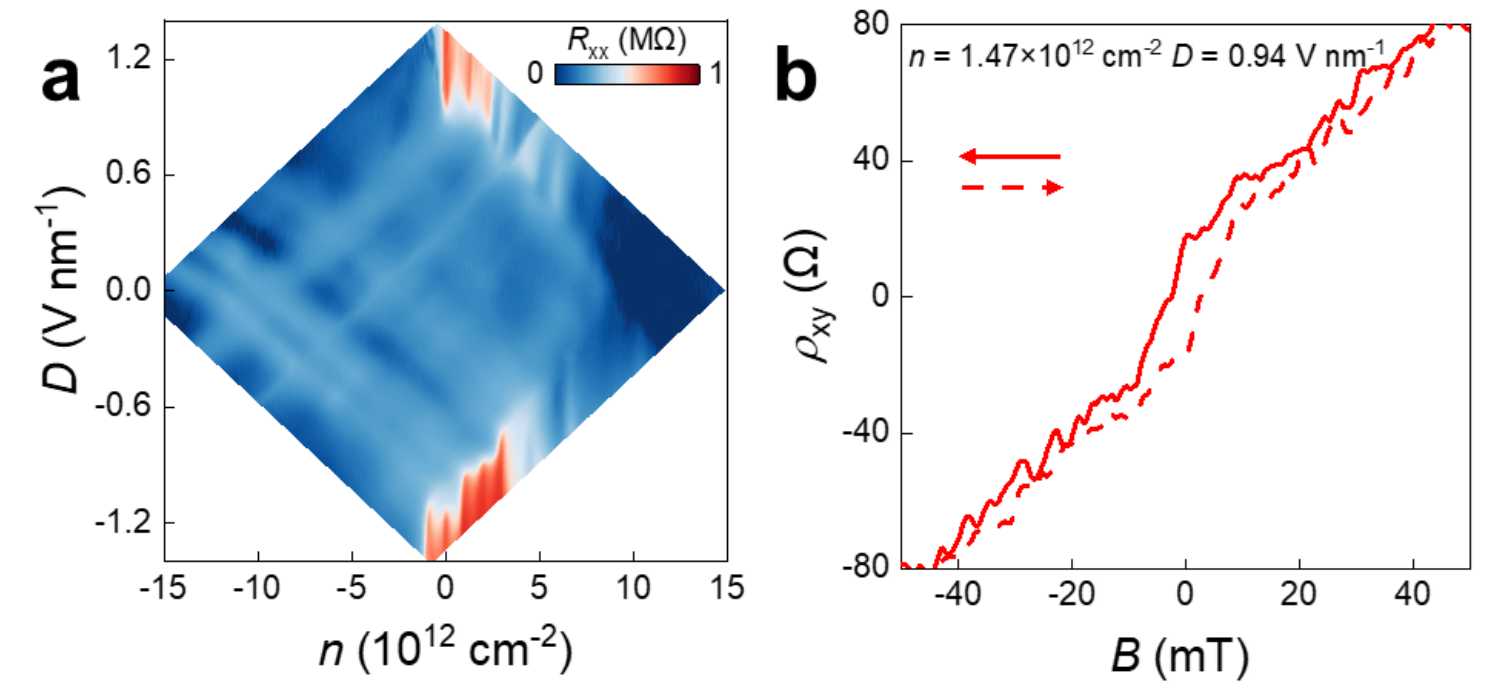

**Extended Data Fig. 11 | Surface ferromagnetic states at a second aligned device (Device D4).** **a,** Color maps of longitudinal resistance $R_{xx}$ as a function of carrier density $n$ and displacement field $D$ measured at $T = 50$ mK and $B = 0$ T. This device shows similar moiré period to Device 2 with almost identical features as those in Fig. 2b, indicating the high homogeneity and reproducibility in graphene/h-BN moiré superlattice. **b**, Anti-symmetrized Hall resistance $\rho_{xy}$ as a function of $B$ swept back and forth at $n = 1.47 \times 10^{12}$ cm$^{-2}$ and $D = 0.94$ V nm$^{-1}$. AHE with both nonlinear features and hysteresis loops was observed at polarized surface states (high $D$ and nonzero $n$), consistent with that in Device D2.

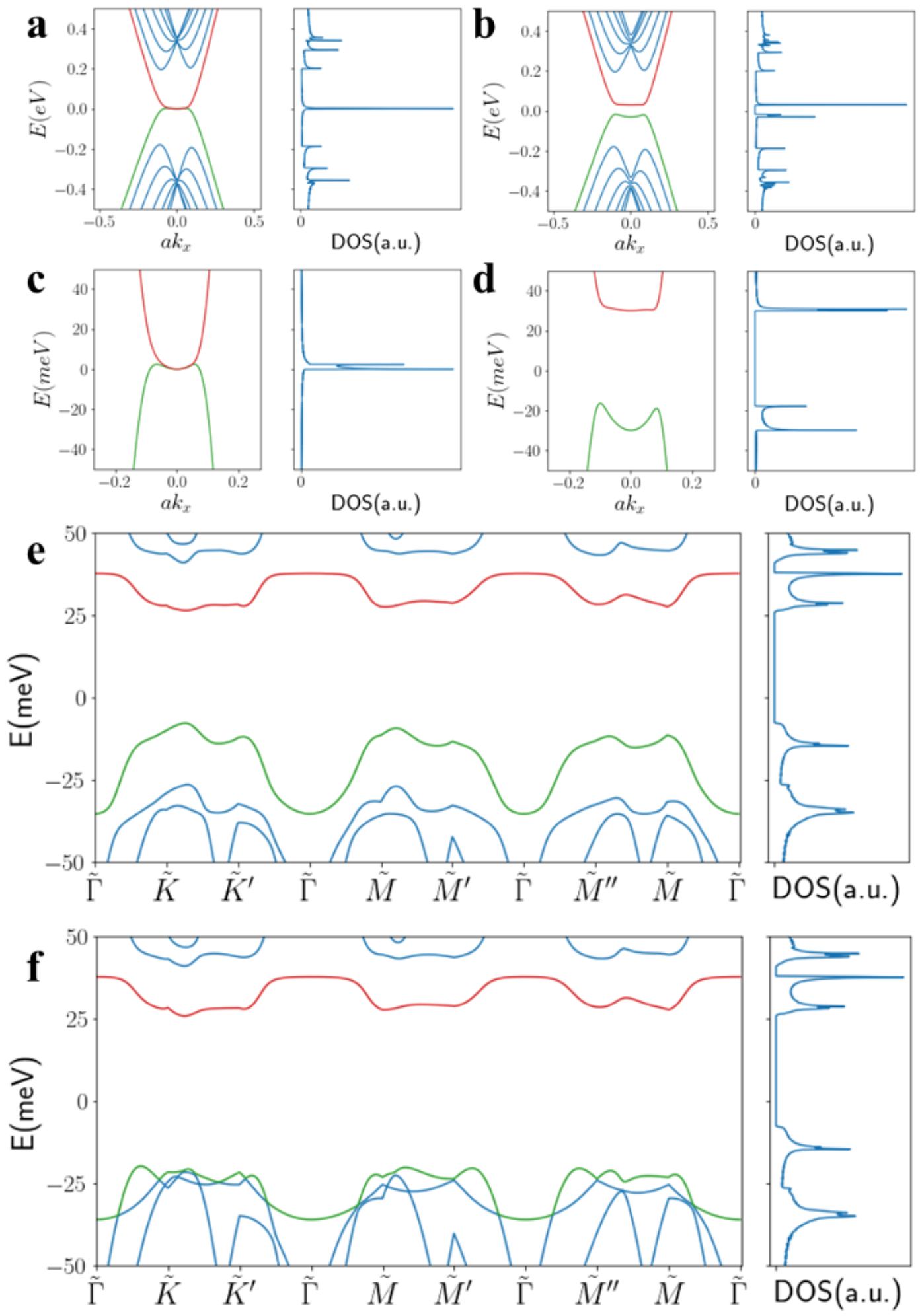

**Extended Data Fig. 12 | Band structure and DOS of rhombohedral 7L graphene. a-d,** Calculated band structure and DOS of intrinsic rhombohedral 7L graphene without moiré superlattice. (**a**) and (**c**) are band structures without interlayer potential. (**b**) and (**d**) are band structures under interlayer potential $\Delta = 10$ mV. (**c**) and (**d**) are the low-energy surface bands. **e**, Calculated band structure of rhombohedral 7L graphene with moiré potential at both top and bottom surface. **f**, Calculated band structure of rhombohedral 7L graphene with moiré potential at only one of the two surfaces. The similar low energy conduction band in (**e**) and (**f**) indicates that the two surfaces are almost decoupled. Right panels in (**a**)-(**f**) are the corresponding DOS as a function of $E$.